\documentclass[11pt]{article}
\usepackage{mathtools}
\usepackage{amsmath, amsthm, amsfonts,amssymb}
\usepackage{enumitem}
\usepackage{graphicx}
\usepackage{colortbl}
\usepackage{tikz}
\usepackage[utf8]{inputenc}
\usepackage{esint}
\usepackage{mathrsfs}
\usepackage[T1]{fontenc}
\usepackage{mathrsfs}
\usepackage{varwidth}
\usepackage{bbm}
\usepackage{enumitem}
\usepackage{mathtools}
\usepackage{array}
\usepackage{graphicx}
\usepackage{caption}
\usepackage{subcaption}
\usepackage{dsfont}
\usepackage{ushort}
\usepackage{accents}
\usepackage{booktabs}

\usepackage[
  backend=biber,   
  style=numeric,   
  giveninits=true  
]{biblatex}

\usepackage{nicefrac}

\usepackage{tikz}
\usetikzlibrary{angles, quotes,calc,patterns}
\usepackage{tikz}
\usepackage{tikz-3dplot}

\usepackage{lineno} 
\usepackage{mathtools}
\numberwithin{equation}{section}
\usepackage{pgfplots}
\pgfplotsset{compat = newest}
\usepackage[titletoc, toc, page]{appendix}

\newcommand{\M}{\mathcal{M}}
\newcommand{\N}{\mathcal{N}}

\newcommand{\eps}{\varepsilon}

\newcommand\blfootnote[1]{%
	\begingroup
	\renewcommand\thefootnote{}\footnote{#1}%
	\addtocounter{footnote}{-1}%
	\endgroup
}
\def\d{\,\mathrm{d}}

\usepackage[colorlinks=true,linkcolor=blue,citecolor=blue,urlcolor=blue,breaklinks]{hyperref}
\usepackage{hyperref}
\usepackage{url}
\usepackage[labelfont=bf]{caption}
\usepackage[toc,page]{appendix}

\def\1{\mathds{1}}

\def\d{\,\mathrm{d}}

\def\p{\,\partial}

\let\eps\varepsilon

\newtheorem{thm}{Theorem}[section]
\newtheorem{cor}[thm]{Corollary}
\newtheorem{lem}[thm]{Lemma}
\newtheorem{prp}[thm]{Proposition}

\newtheorem{dfn}[thm]{Definition}
\newtheorem{remark}[thm]{Remark}

\hypersetup{pdftitle={Anisotropic}}
\hypersetup{pdfauthor={Merino, Moschella}}
\author{Sara Merino-Aceituno\footnote{Faculty of Mathematics, University of Vienna, Oskar-Morgenstern-Platz 1, 1090 Vienna, Austria. \\ sara.merino@univie.ac.at \& carmela.moschella@univie.ac.at}
\and
Carmela Moschella\footnote{
Mathematical Institute, University of Oxford,
Andrew Wiles Building, Radcliffe Observatory Quarter,
Woodstock Road, Oxford OX2 6GG, UK.\\
carmela.moschella@maths.ox.ac.uk
} {{$^*$}}
}
\title{The Vicsek--Kuramoto model in collective dynamics: macroscopic equations and pattern formation}
\makeindex

\begin{document}
\maketitle
	
	\vspace{-10pt}

	\begin{abstract}
		\noindent
In this work, we investigate an individual-based model (IBM) for self-propelled agents interacting locally on a plane. Agents are characterized by their position, the angle determining their direction of motion, and their angular velocity. The dynamics combine features of the well-known Vicsek and Kuramoto models, which describe collective dynamics and synchronization, respectively. The evolution of the directions of motion follows a Vicsek model, where agents align their orientations with the mean orientation of their neighbors, subject to some noise. Similarly, the angular velocities relax towards the average angular velocity of the neighboring agents, also subject to noise.

From the IBM we derive the corresponding kinetic equation in the limit of a large number of agents and formally obtain the macroscopic equations through a macroscopic (hydrodynamic) limit. Numerical simulations of the IBM reveal a variety of patterns, including rotating clusters, traveling orientation waves, and globally synchronized rotational motion. A qualitative comparison with simulations of the macroscopic system show the ability of the macroscopic model to reproduce some emergent behavior of the IBM.

		\blfootnote{\emph{Keywords and phrases.} mean-field limit, continuum equations, kinetic equations}
		\blfootnote{\emph{2020 Mathematics Subject Classification.} 35Q92, 82C22, 82D30, 82B40}
	\end{abstract}
	
	\tableofcontents

\section{Introduction} \label{sec:intro}

\subsection{Vicsek, Kuramoto and related models}

This work investigates an individual-based model (IBM) for the collective behavior of self-propelled agents subject to multiple local alignment mechanisms. We consider a third-order particle system in which the angular velocity evolves in response to interactions. Specifically, agents move on a plane and are described by their position, the angle specifying their direction of motion, and their angular velocity. The dynamics merge features of the Vicsek and Kuramoto models, capturing both collective motion and synchronization; we therefore refer to this system as the Vicsek--Kuramoto (VK) model. Agents update their orientation via a Vicsek-like rule: each aligns its direction of motion with the local mean orientation under stochastic perturbations. Likewise, angular velocities relax toward the neighbors’ mean angular velocity, also subject to noise.

Both the Vicsek and Kuramoto models have played a fundamental role in modeling collective behavior, particularly in biological systems. The Vicsek model, initially proposed as a minimal framework for flocking, exhibits rich emergent dynamics, including a phase transition between disordered and ordered motion~\cite{aldana2007phase,chate2008collective,degond2013phase,degond2020phase,merino2025stability,paley2007stability,vicsek1995novel}, traveling density bands~\cite{ginelli2016physics}, and milling states~\cite{liebchen2016rotating}. In the many-particle limit, the system is approximated by a mean-field kinetic equation governing the time evolution of the distribution of individuals in position and velocity~\cite{BolleyCanizoCarrillo2012,bolley2011mean,briant2022cauchy,degondcontinuum}. Further coarse-graining yields macroscopic equations for macroscopic observables such as particle density and mean orientation. The first macroscopic description, Self-Organized Hydrodynamics (SOH), was proposed in~\cite{degondcontinuum} and later made rigorous in~\cite{jiang2016hydrodynamic}.

There is also a vast literature on the Kuramoto model. The original Kuramoto system is a first-order model of synchronization in coupled oscillators. Depending on system parameters, one observes phase transitions from incoherence to partially or fully synchronized states~\cite{ha2010complete,ha2016emergence,ha2016synchronization}. Mean-field and macroscopic descriptions have likewise been derived for this system~\cite{choi2018emergent,HaLiu2009,ha2008particle}.

Beyond the intrinsic interest of combining two well-studied models, our work is motivated by applications to real systems. The proposed model is closely related to those in~\cite{weaksync} from the physics literature, which successfully explain patterns observed in bacterial suspensions, and to the model in~\cite{sumino2012vortex}. These models account for rotational behaviors such as vortex formation in dense suspensions of self-propelled agents. Similar patterns have been observed in other biological systems, for example in suspensions of sperm cells forming stable vortex arrays~\cite{riedel2005self}. Moreover, our work extends the theoretical framework of~\cite{kuramoto-vicsek}, where a related Kuramoto--Vicsek model was introduced. Unlike our setting, agents in~\cite{kuramoto-vicsek} are assigned a constant angular velocity, leading to persistent circular trajectories. In contrast, our model introduces third-order dynamics in which angular velocity evolves in time and is influenced by interactions with neighboring agents. Although~\cite{kuramoto-vicsek} develops a related theory, it does not provide numerical results at either the microscopic or macroscopic scale.

It is also worth noting that our model bears similarities to others in the literature, such as the Persistent Turning Walker with Alignment (PTWA) and the swarmalator model. The PTWA model~\cite{ptwa} is a third-order particle system in which each agent moves at a constant speed and updates its orientation via the curvature of its trajectory, which relaxes toward a target curvature. This target curvature is computed from the misalignment between the agent’s direction of motion and the mean direction of its neighbors within a fixed interaction radius. As observed, for example, in~\cite{MemoryEffect}, this framework captures key aspects of persistent trajectories and curvature-based coordination characteristic of schooling fish. Indeed, underdamped dynamics, where the effective force acts on a particle’s curvature rather than directly on its direction of motion, have been shown to better capture the smooth directional changes characteristic of chemically propelled rods~\cite{Paxton2004Catalytic,Takagi2013Dispersion}, steadily swimming fish~\cite{gautrais2009analyzing}, and biofilaments gliding in high motor density motility assays~\cite{sumino2012vortex}.
Finally, another related model is the swarmalator~\cite{degond2022topological}, in which each particle moves at a constant self-propulsion speed and carries an internal phase evolving under a combination of interaction mechanisms: alignment of propulsion direction toward local averages, synchronization of internal phases, attraction--repulsion forces, and external confinement. Notably, the microscopic simulations in~\cite{degond2022topological} exhibit traveling-wave patterns in the phase variable that qualitatively resemble those arising in our model under suitable parameter regimes. Although the two models display similar behaviors, their underlying dynamics differ fundamentally. The swarmalator model features non-symmetric interactions, with each agent’s phase directly influencing its spatial motion. In contrast, our model employs symmetric interactions, and angular velocity affects motion only indirectly, through its influence on orientation.

\subsection{Scope and aims}

The first aim of the present work is to derive a mean-field description of the  Vicsek--Kuramoto model presented in section \ref{sec:IBM}. In the limit of a large number of interacting particles, this leads to a kinetic equation of Fokker–Planck type for the one-particle distribution function, defined over position, orientation, and angular velocity space.

Next, we seek a macroscopic description of the dynamics via a hydrodynamic (macroscopic) limit. A common feature of active particle systems, such as the Vicsek model, is that momentum and kinetic energy are generally not conserved. As a result, the system lacks a sufficient number of standard conserved quantities, or collisional invariants, to close the macroscopic equations. To overcome this difficulty, we use the framework of Generalized Collisional Invariants \cite{degondcontinuum}, which allows us to derive a closed macroscopic system describing the agent density, mean orientation, and average angular velocity.
The resulting equations form an Euler-type system, consisting of two conservative equations for the mass and the angular momentum, coupled with an evolution equation for the mean orientation.

Our motivation for deriving macroscopic equations is that macroscopic dynamics provide the natural scale at which group-level phenomena and emergent behaviors, such as pattern formation, can be observed. By linking microscopic and macroscopic descriptions, we connect individual behavior with collective behaviour. However, some important information from the particle dynamics may be lost in this coarse-graining. To assess the validity of the macroscopic model and its ability to approximate particle dynamics, we perform numerical simulations to qualitatively compare its behavior with that of the IBM. Simulations of the IBM exhibit various collective behaviors, including rotating clusters, traveling orientation waves, and synchronized oscillatory states. We then compare these solutions to those of the macroscopic system.

\subsection{Structure of the paper}
 
The paper is organized as follows. In Section \ref{sec:IBM_mean_field}, we introduce the individual-based model, derive its mean-field kinetic description, and present the hydrodynamic scaling. Section \ref{sec:main_result} states the main theorem of the paper and details the derivation of the macroscopic equations using the Generalized Collision Invariant (GCI) method. Section \ref{sec:simulation_micro} presents numerical simulations of the IBM, highlighting the emergence of various collective behaviors. Section \ref{sec:simulation_macro} introduces the macroscopic simulations and provide a qualitative comparison with those of the IBM. Appendix \ref{appendix:videos} contains supplementary videos illustrating representative microscopic and macroscopic simulation results discussed in Sections \ref{sec:simulation_micro} and \ref{sec:simulation_macro}.

\section{Individual-Based model, mean-field limit and scaling}\label{sec:IBM_mean_field}
\subsection{The IBM}\label{sec:IBM}
We consider a system of $N$ interacting agents moving in the two-dimensional plane $\mathbb{R}^2$ at constant speed $c > 0$. The state of the $i$-th agent at time $t$ is described by its position $x_i(t) \in \mathbb{R}^2$, orientation angle $\theta_i(t) \in (-\pi, \pi]$ and angular velocity $\omega_i(t) \in \mathbb{R}$. The unit vector $
\vec{\tau}(\theta_i) = (\cos \theta_i, \sin \theta_i)
$ gives the direction of motion.
The following system of coupled stochastic differential equations governs the dynamics of the system
\begin{align}
\label{eq:VicsekKuramoto1}
\frac{dx_i}{dt}&= c \, \Vec{\tau}(\theta_i), \\
\label{eq:VicsekKuramoto2}d\theta_i &= \omega_i \,dt + k_{\theta}\sin(\overline{\theta}_i- \theta_i)\, dt + a\, dB_t^i, \\ 
\label{eq:VicsekKuramoto3}d\omega_i &= k_{\omega}(\overline{\omega}_i- \omega_i) dt + b\, d\tilde B_t^i,  
\end{align}
where $B^i_t, \tilde B^i_t$ are standard Brownian motions, independent between themselves and those of each agent $i \neq j$. The positive constants $a$ and $b$ quantify the intensity of the random perturbation acting on the orientation angle and the angular velocity, respectively. In this model, agents are subject to two alignment mechanisms. The first induces a relaxation of the orientation angle $\theta_i$ of each agent towards the average orientation of its neighbors $\bar\theta_i$, with a relaxation frequency $k_\theta >0$. The second prescribes a relaxation of the angular velocity $\omega_i$ of each agent towards the local average angular velocity $\bar{\omega}_i$, with relaxation frequency $k_\omega>0$. In the absence of these interaction forces and the Brownian motions, each particle is subjected to its intrinsic angular velocity $\omega_i$, which results in a circular trajectory of radius $r_i = \frac{c}{|\omega_i|}$, rotating counter-clockwise if $\omega_i > 0$ and clockwise if $\omega_i < 0$.
The average direction $\bar{\theta}_i$ is defined as the orientation corresponding to the normalized local average of the velocity directions of surrounding agents, where the influence of each neighbor is modulated by a radially symmetric interaction kernel \( K = K(|x|) \), i.e. 
\begin{equation}
 \Vec{\tau}(\overline{\theta}_i) = \overline{\Omega}_i\,,\qquad \overline{\Omega}_i = \frac{{J}_i}{|{J}_i|}\,, \qquad 
{J}_i = \frac{1}{N} \sum_{\begin{subarray}{l}j = 1\\ 
\end{subarray}}^ N K\big({|x_i-x_j|}\big)\,{\Vec{\tau}}(\theta_j).
\end{equation}
Similarly, $\bar{\omega}_i$ denotes the local average angular velocity among neighboring agents
\begin{equation}
\begin{aligned}
\bar{\omega}_i =\frac{1}{\mathcal{N}_i} \sum_{\begin{subarray}{l}j = 1\\ 
\end{subarray}}^ N K\left({|x_i-x_j|}\right)\omega_j\,,\qquad \mathcal{N}_i = {\sum_{j=1}^N K\left( {|x_i - x_j|} \right)}\,,
\label{eq:bar_omega}
\end{aligned}
\end{equation}
where $\mathcal{N}_i$ indicates the number of neighboring particles weighted by the kernel $K$. Moreover we impose that $$\int K dx =1.$$ We observe that the interaction term for $\omega_i$ is divided by $\mathcal{N}_i$. This normalization ensures that the interaction force remains of order one, regardless of the total number of particles. In contrast, such normalization is not required in the orientation interaction term, as the vector $\Omega_i$ is of order one by construction.

 As a preliminary step toward deriving the macroscopic model, we introduce a non-dimensional form of the system \eqref{eq:VicsekKuramoto1}--\eqref{eq:VicsekKuramoto3}. We choose a fixed time scale $ t_0 $ and the associated space scale $ x_0 = c \,t_0 $, which represents the typical distance traveled by a particle over that timescale. The characteristic angular velocity scale is set to $\omega_0 = t_0^{-1} $. Moreover,  \( \overline{\theta}_i \) remains unchanged since it represents a direction on the unit circle, while \( \overline{\omega}_i \) rescales as \( \tilde{\overline{\omega}}_i = \frac{\overline{\omega}_i}{\omega_0} \), consistently with the scaling of \( \omega_i \).
With these choices, we introduce the following dimensionless variables $\Tilde{x}:= \frac{x}{x_0}$, $\Tilde{t}:= \frac{t}{t_0}$ and $\Tilde{\omega}:= \frac{\omega}{\omega_0}$ and the rescaled interaction kernel \( \tilde K \), defined by \( K(x_0 |\tilde x|) = \tilde K(|\tilde{x}|) \). In particular, if \( K \) is the indicator function of a ball of radius \( R \), then \( \tilde K \) corresponds to the indicator function of a ball of radius \( \tilde R = R / x_0 \).  The noise intensities are now encoded in the dimensionless parameter \( \alpha^2 = \frac{a^2 t_0}{2} \,\text{and}\,\beta^2 = \frac{b^2 t_0}{2} \), while the rescaled interaction forces are given by
\[
\tilde k_{\theta} = { k_{\theta}}t_0\,, \qquad \tilde k_{\omega} = { k_{\omega}}t_0\,.
\] This leads to the following dimensionless system (dropping the tildes for simplicity):
\begin{eqnarray}
dx_i &=& \Vec{\tau}(\theta_i) \, dt\,, \label{eq:VicsekKuramoto_dimensionless1}\\
d\theta_i &=& \omega_i \, dt +  k_\theta \sin(\overline{\theta}_i - \theta_i) \, dt +\sqrt{2\alpha^2} dB_t^i\,, \label{eq:VicsekKuramoto_dimensionless2}\\
d\omega_i &=&  k_\omega (\overline{\omega}_i - \omega_i) \, dt + \sqrt{2\beta^2} \, d\tilde B_t^i\,. \label{eq:VicsekKuramoto_dimensionless3}
\end{eqnarray}

\subsection{Mean--field equations}
The first step in our analysis consists in deriving the mean-field equation corresponding to the IBM \eqref{eq:VicsekKuramoto_dimensionless1}--\eqref{eq:VicsekKuramoto_dimensionless3}.

As is standard in the mean-field analysis of interacting particle systems (see, e.g., \cite{spohn2012large}), we study the empirical distribution associated with the particle system,
\begin{equation}\label{eq:empirical_distr}
    f^N(t, x, \theta, \omega) = \frac{1}{N} \sum_{i=1}^N \delta_{(x_i(t), \theta_i(t), \omega_i(t))}(x, \theta, \omega)\,.
\end{equation}
Formally, in the limit for $N\to \infty$ the empirical distribution converges to the distribution function $f$ that satisfies the following Fokker-Planck equation
\begin{equation}\label{eq:kientic_model}
    \partial_t\,f + \Vec{\tau}(\theta) \cdot \nabla_x f  + \partial_{\theta}(\omega\, f) = -k_{\theta}\,\partial_{\theta}(f\,F^K_f) - k_{\omega}\,\partial_{\omega}(f\,G^K_f) + \alpha^2\partial_{\theta}^2f + \beta^2\partial_{\omega}^2f\,,
\end{equation}
with
\begin{equation}\label{def:F_alignment}
\begin{aligned}
       &F^K_{f} = \sin(\overline{\theta}_{f}-\theta)\,, \quad  \Vec{\tau}(\overline{\theta}_{f})(t,x)= \overline{\Omega}^K_{f}\,,\quad \overline{\Omega}^K_{f}= \frac{{J}^K_{f}({x})}{|{J}^K_{f}({x})|}\,,\\
       & {J}^K_{f}=\int_{\mathbb{R}^2\times(-\pi, \pi]\times\mathbb{R}} K_R\big ({|x-y|} \big)\Vec{\tau}(\theta)f(t,y,\theta,\omega) dy\,d\theta\, d\omega\,,
       \end{aligned}
\end{equation}
\normalsize
\begin{equation}\label{def:G_alignment}
\begin{aligned}
       G^K_{f} = \overline{\omega}_{f}-\omega\,,\quad \overline{\omega}_{f}(t,x)&= \frac{1}{\rho^K_f(x)}\int_{\mathbb{R}^2\times(-\pi, \pi]\times\mathbb{R}}K_R\big({|x-y|} \big)\omega\,f(t,y,\theta,\omega) dy\, d\theta\, d\omega\,,  \\
       \rho^K_f(x)&= \int_{\mathbb{R}^2\times(-\pi, \pi]\times\mathbb{R}} K_R\big ( {|x-y|}\big)\,f(t,y,\theta,\omega) dy\, d\theta\, d\omega\,,
\end{aligned}      
\end{equation}
where $K_R$ is the scaled interaction kernel 
\[
K_R(x) := \frac{1}{R^2} K\left( \frac{x}{R} \right) 
\]
so that
\[
\int_{\mathbb{R}^2} K_R  (x)\, dx = 1,
\]
where the parameter $R>0$ represents the typical interaction radius.

The proof of the mean--field limit can be made rigorous since all terms are Lipschitz, following classical coupling arguments (see for example \cite{BolleyCanizoCarrillo2012, sznitman1991topics, carmona2016lectures}).

\subsection{Scaling}\label{sec:scaling}
We now make the following scaling assumptions, for a scaling parameter $\varepsilon \ll 1$:
\begin{equation}\label{def:scaling}
R = \varepsilon\,, \quad k_\theta = \mathcal{O}\left(\frac{1}{\varepsilon}\right), \quad k_\omega = \mathcal{O}\left(\frac{1}{\varepsilon}\right), \quad \beta^2 = \mathcal{O}\left(\frac{1}{\varepsilon}\right), \quad \alpha^2 = \mathcal{O}\left(\frac{1}{\varepsilon}\right).   
\end{equation}
This scaling reflects a regime in which interaction forces and stochastic fluctuations dominate the dynamics.
 Thus, introducing $k'_\theta$, $k'_\omega$, $\beta'$, and $\alpha'$ such that $k_\theta = k'_\theta/\varepsilon$, $k_\omega = k'_\omega/\varepsilon$, $\beta^2 = (\beta')^2/\varepsilon$, and $\alpha^2 = (\alpha')^2/\varepsilon$, we may assume that $k'_\theta$, $k'_\theta$, $\beta'$ and $\alpha'$ are constants. After this scaling, equation \eqref{eq:kientic_model} is written as (dropping the primes for simplicity):
\begin{equation}\label{main1}
    \p_t f^\eps + \Vec{\tau}(\theta) \cdot \nabla_x f^\eps + \partial_{\theta}(\omega\,f^\eps) = \frac{1}{\varepsilon} Q(f^\eps)\,,
\end{equation}
where 
\begin{equation}
Q(f^\eps) = 
k_{\theta}\,\p_{\theta}(F^K_{f^\eps}\,f^\varepsilon) + \alpha^2 \partial_{\theta}^2 f^\eps + k_{\omega}\,\partial_{\omega}(G^K_{f^\varepsilon}\,f^\eps)+ \beta^2\partial_{\omega}^2\,f^\eps\,.
\end{equation}

We now state the following lemma, which characterizes the localization of the interaction terms in space.

\begin{lem}[Expansion for localized interactions]\label{lem:expansion}
The following expansions hold:
\begin{equation*}
    \begin{aligned}
        F^K_{f^\varepsilon} &= F_{f^\varepsilon} + \mathcal{O}(\varepsilon^2), \\
        G^K_{f^\varepsilon} &= G_{f^\varepsilon} + \mathcal{O}(\varepsilon^2),
    \end{aligned}
\end{equation*}
where the leading-order terms are given by
\begin{equation*}
\begin{aligned}
&F_f = \sin(\overline{\theta}_f - \theta)\,, 
\quad 
\Vec{\tau}(\overline{\theta}_f)(t,x) = \Omega_f(t,x),
\quad 
\Omega_f = \frac{{J}_f(x)}{|{J}_f(x)|}\,,\\
&{J}_f(x) = \int_{(-\pi, \pi] \times \mathbb{R}} \Vec{\tau}(\theta)\, f(t,x,\theta,\omega)\, d\theta\, d\omega\,,\\
G_f = \overline{\omega}_f - \omega,
\quad 
\overline{\omega}_f(t,x) &= \frac{1}{\rho_f(x)}\int_{(-\pi, \pi] \times \mathbb{R}} \omega\, f(t,x,\theta,v)\, d\theta\, d\omega\,, \quad
       \rho_f(x) = \int_{(-\pi, \pi]\times\mathbb{R}}\,f (t,y,\theta,\omega)\, d\theta\, d\omega\,.
\end{aligned}
\end{equation*}
\noindent
\end{lem}
\begin{proof}
We remind that the scaled interaction kernel is defined as
\[
K_\varepsilon(x) := \frac{1}{\varepsilon^2} K\left( \frac{|x|}{\varepsilon} \right)\,,
\]
hence we can rewrite the nonlocal alignment term as
\begin{equation}
\begin{aligned}
{J}_f(t,x) &= \int_{\mathbb{R}^2 \times (-\pi, \pi] \times \mathbb{R}}\frac{1}{  \varepsilon^2}  K\left( \frac{|x - y|}{\varepsilon} \right) \Vec{\tau}(\theta)\, f(t, y, \theta, \omega) \, dy\,d\theta\,d\omega\\
&=  \int_{\mathbb{R}^2 \times (-\pi, \pi] \times \mathbb{R}} K( |z|) \,\Vec{\tau}(\theta)\, f(t, x - \varepsilon z, \theta, \omega) \, dz\,d\theta\,d\omega\,,
\end{aligned} 
\end{equation}
where in the last equality we have used the change of variable $y = x - \varepsilon z$, so $dy = \varepsilon^2 dz$.
Assuming that $f$ is smooth enough in the spatial variable, we perform a Taylor expansion at $x$:
\begin{equation} \label{eq:Taylor_expansion}
f(t, x - \varepsilon z, \theta, \omega) = f(t, x, \theta, \omega) - \varepsilon z \cdot \nabla_x f(t, x, \theta, \omega) + \frac{ }{2} z^T \nabla_x^2 f(t, x, \theta, \omega) z + \mathcal{O}(\varepsilon^3)\,,
\end{equation}
and we substitute it in the previous expressions, where $z^T$ denotes the transpose of $z$. Finally, using the fact that $K$ is even, i.e.
\[
\int_{\mathbb{R}^2} z\, K\left( \frac{|z|}{r'} \right) dz = 0,
\]
we find the leading-order approximation
\begin{equation}
{J}_f(t,x) 
= \int_{(-\pi, \pi] \times \mathbb{R}} \Vec{\tau}(\theta)\, f(t,x,\theta,\omega)\, d\theta\,d\omega 
+ \mathcal{O}(\varepsilon^2).
\end{equation}
A similar argument applies to $G^K_{f^\varepsilon}$, concluding the proof.
\end{proof}

The expansion in the previous lemma leads to the following expansion of the kinetic equation:
\begin{equation}\label{eq:kinetic_scaled1}
    \p_t f^\eps + \Vec{\tau}(\theta) \cdot \nabla_x f^\eps + \partial_{\theta}(\omega\,f^\eps) = \frac{1}{\varepsilon} Q(f^\eps) + \mathcal{O}(\varepsilon^2) \,,
\end{equation}
where 
\begin{equation}\label{eq:collision_operator}
Q(f ) = 
k_{\theta}\,\partial_{\theta}(f\,F_f) + k_{\omega}\,\partial_{\omega}(f\,G_f) + \alpha^2\partial_{\theta}^2f + \beta^2\partial_{\omega}^2f\,,
\end{equation}
with $F_f$ and $G_f$ given by Lemma~\ref{lem:expansion}.
The hydrodynamic model is obtained as the $\varepsilon\to 0$ limit of this system.

\section{Macroscopic limit}\label{sec:main_result}

\subsection{Main result}
In this section, we state the main theoretical result of our analysis, namely the derivation of the macroscopic system obtained as 
the formal limit for $\varepsilon\to 0$ of equation \eqref{eq:kinetic_scaled1}. To this end, we begin by introducing the following distributions:
\begin{equation} \begin{aligned}\label{eq:equilibria}
    &\M_{\overline{w}_f}(\omega)= C_1\, \exp \Bigg( -\frac{k_{\omega}}{2\beta^2}( {\omega-\overline{\omega}_{f}})^2 \Bigg)\,, \qquad C_1=\sqrt{{k_\omega}/{2\pi \beta^2}}\,,\\
    &\N_{\overline{\theta}_f}(\theta)= C_2\, \exp \Bigg( \frac{k_{\theta}}{\alpha^2} \cos(\theta-\overline{\theta}_{f})\Bigg)\,,\qquad C_2
    = \frac{1}{2\pi I_0\left({k_\theta}/{\alpha^2}\right)}\,,
    \end{aligned}
\end{equation}
where \(\M_{\overline{w}_f}\) corresponds to a Gaussian distribution centered at \(\overline{\omega}_f\) and with variance $\beta^2/k_\omega$. \(\mathcal{N}_{\overline{\theta}_f}(\theta)\) is the von Mises distribution on the interval \((-\pi, \pi]\) centered at \(\overline{\theta}_f\) and with variance $1-\frac{I_1(k_\theta/\alpha^2)}{I_0(k_\theta/\alpha^2)}$, where
\( I_j(\cdot) \) denotes the modified Bessel function of order $j$.
We now proceed to the statement of the main theorem. 
\begin{thm}\label{thm:main_macro}
    Suppose that there is a smooth solution $f^\varepsilon$ to the kinetic model \eqref{eq:kinetic_scaled1} for all $\varepsilon>0$ and that this solution converges as $\varepsilon\to 0$ to some function $f^0$ strongly enough so that we can exchange limits with integrals and derivatives. Then 
    \begin{equation}\label{eq:equilibria_distr}
    f^\varepsilon \xrightarrow[\varepsilon \to 0]{} f^0 = \rho\, \mathcal{N}_{\overline{\theta}}(\theta)\, \M_{\overline{w}}(\omega)\,,     \end{equation}
where \(\mathcal{N}_{\bar\theta}\) and \(\mathcal{M}_{\bar\omega}\) are respectively the von Mises distribution and the Gaussian defined in \eqref{eq:equilibria}. The density $\rho = \rho(t,x)$, the mean orientation $\Omega= \Omega(t,x)= (\cos \bar \theta(t,x), \sin \bar \theta(t,x))$  and the mean angular velocity $\bar \omega = \bar \omega (t,x)$ satisfy
the following macroscopic system:
\begin{align}
&\partial_t \rho + c_1 \nabla_x \cdot (\rho \, \Omega) = 0\,, \label{eq:macro_eq1}\\
&\partial_t (\rho \, \overline{\omega}) + c_1 \nabla_x \cdot (\rho \, \overline{\omega} \, \Omega) = 0\,, \label{eq:macro_eq2}\\
&\rho \left(  \partial_t \Omega 
+ c_2 (\Omega \cdot \nabla_x) \Omega 
- \, \overline{\omega} \, \Omega^\perp \right)
+ \frac{1}{\kappa} P_{\Omega^\perp} \nabla_x \rho = 0\,, \label{eq:macro_eq3}
\end{align}
where \(P_{\Omega^\perp} = \mathrm{Id} - \Omega \otimes \Omega\) is the projection onto the orthogonal direction \(\Omega^\perp = (-\sin \overline{\theta}, \cos \overline{\theta})\), with \( \mathrm{Id} \) denoting the \(2 \times 2\) identity matrix. The constant $c_1$ is defined as \begin{equation}\label{def:c1}
    c_1 = \frac{I_1\left( \kappa \right)}{I_0\left( \kappa\right)}\,,
\end{equation} where $\kappa = \frac{k_\theta}{\alpha^2}$, and $c_2=K_2/K_1$ with
\begin{equation}\label{def:K1}
 K_1 := \int \sin(\theta)\, \mathcal{N}_0(\theta)\, g(\theta)\, d\theta\,,   \qquad K_2 := \int \cos(\theta)\, \sin(\theta)\, \mathcal{N}_0(\theta)\, g(\theta)\, d\theta.
\end{equation}

\end{thm}
The macroscopic system \eqref{eq:macro_eq1}--\eqref{eq:macro_eq3} shares structural similarities with those derived in previous works on alignment-based models such as ~\cite{degondcontinuum, ptwa}. In particular, it has the same form as the macroscopic system obtained in~\cite{kuramoto-vicsek} in the regime of small angular velocity (SOHR-S model), despite being derived from a different microscopic dynamics. System~\eqref{eq:macro_eq1}--\eqref{eq:macro_eq2} consists of two conservative equations, respectively for the mass density $\rho$ and the angular momentum density $\rho \,\bar\omega$. In particular as observed in \cite{kuramoto-vicsek}, the equation for $\rho \,\bar\omega$ can be rewritten  as a transport equation for the average rotation velocity $\bar\omega$ using equation \eqref{eq:macro_eq1}:
\begin{equation}
    \partial_t \bar\omega + c_1 \Omega \cdot \nabla_x \bar\omega = 0,
    \label{eq:transport_Y}
\end{equation}
which simply expresses that the average angular velocity $\bar\omega$ is convected at speed $c_1$ along the direction $\Omega$.
The equation~\eqref{eq:macro_eq3} governs the evolution of the average direction of motion $\Omega$, subject to the geometrical constraint $|\Omega| = 1$. This constraint is dynamically preserved by the presence of the projection matrix 
$
P_{\Omega^\perp} = \mathrm{Id} - \Omega \otimes \Omega\,.
$ Indeed by taking the scalar product of the equation with $\Omega$, we obtain $
\partial_t |\Omega|^2 + c_2 (\Omega \cdot \nabla_x) |\Omega|^2 = 0\,,$
which shows that the norm $|\Omega|$ remains constant along the flow. Owing to the geometrical constraint $|\Omega|=1$, the equation for 
$\Omega$ cannot be written in conservative form, as the dynamics evolve on the unit circle. This intrinsic non-conservative structure reflects the absence of momentum conservation in the underlying microscopic model.
Finally, a key feature of this equation is the presence of the term $-\overline{\omega}\,\Omega^\perp$, which describes the turning of the collective direction field in proportion to the average angular velocity $\overline{\omega}$, encoding the influence of individual self-rotation at the macroscopic scale.

\begin{remark}
    Notably, the parameters $k_\omega$ and $\beta^2$ do not appear explicitly in the macroscopic system \eqref{eq:macro_eq1}--\eqref{eq:macro_eq2}. They enter only through the limiting distribution \eqref{eq:equilibria_distr}, via their ratio $k_\omega/\beta^2$ in the Gaussian $\mathcal{M}_{\bar{\omega}}$. This reduction already suggests important consequences for the patterns the macroscopic model can reproduce since by eliminating an independent parameter, the system loses one degree of freedom.
\end{remark}

\subsection{Investigation of particular solutions}

Notice that constant values for $(\rho, \Omega, \bar \omega)$ is not a solution of \eqref{eq:macro_eq1}-\eqref{eq:macro_eq3} unless $\bar\omega=0$. In this sections we explore different solutions by assuming that one or more of the macroscopic quantities are constant. 

\subsubsection{Solutions with constant angular velocity}

When $\bar\omega$ is constant, then we recover the equations from Ref. \cite{kuramoto-vicsek}. Notice that if $\bar \omega$ is constant but different from zero, then necessarily $\rho$ and $\Omega$ cannot be constant at the same time, or the equation for $\Omega$ \eqref{eq:macro_eq3} cannot be satisfied. If $\bar\omega\equiv 0$, then we recover the SOH model -- the macroscopic model for the Vicsek model \cite{degondcontinuum}.

\subsubsection{Solutions with constant orientation $\Omega$} 

\begin{lem}
   If $\Omega=\Omega_0$ is constant, then $(\rho, \Omega_0,\bar\omega)$ is a solution of \eqref{eq:macro_eq1}-\eqref{eq:macro_eq3} if and only if the initial conditions $\rho_0=\rho_0(x)$ and $\bar\omega_0=\bar\omega_0(x)$ for $\rho$ and $\bar\omega$ satisfy:
   \begin{align} \label{eq:aux_lem_constraint}
   \rho_0\bar\omega_0=\frac{1}{\kappa}\nabla_x\rho_0 \cdot\Omega_0^\perp,
   \end{align}
   and
   \begin{align}
   \rho(t,x)=\rho_0(x-c_1\Omega_0t), \quad \bar\omega(t,x)=\bar\omega_0(x-c_1\Omega_0t), \qquad \Omega=\Omega_0. \label{eq:aux_lem_sol}
    \end{align}
\end{lem}
\begin{proof}
If $\Omega=\Omega_0$ is constant, then the equations for $\rho$ and $\bar\omega$ become transport equations with speed $c_1\Omega_0$ and their solutions are given in \eqref{eq:aux_lem_sol}. 
Now, imposing that $\Omega=\Omega_0$ is constant in the equation for $\Omega$ \eqref{eq:macro_eq3} we obtain the constraint \eqref{eq:aux_lem_constraint}.
\end{proof}

\medskip
Therefore solutions with constant orientation exist, but they require very well prepared initial data. For example:
$$\rho_0=\exp(-|x|^2),\quad \omega_0=-\frac{2}{\kappa}(x\cdot\Omega_0^\perp),$$
or
$$\rho_0=\sin(x_1)+2,\quad \omega_0=-\frac{1}{\kappa (\sin(x_1)+2)}\ (\cos(x_1),0)\cdot\Omega_0^\perp.$$

As we will see later, this type of special solutions are not particularly interesting when looking at the behaviour of the IBM, so we will not explore them further.

\subsubsection{Solutions with synchronised behaviour or constant initial data}

When a solution to the system \eqref{eq:macro_eq1}--\eqref{eq:macro_eq3} is such that $\Omega$ is space-independent and time-periodic, we call it a solution with `synchronised behaviour'.     A particular instance of this is given in the following:
\begin{prp} \label{prop:solutions_constant_density}
    Suppose that there exists a solution $(\rho, \Omega, \bar \omega)$ to \eqref{eq:macro_eq1}-\eqref{eq:macro_eq3} with $\rho=\rho_0$ constant, $\bar\omega=\omega_0$ constant and $\Omega=\Omega(t)$ space independent. Then, $\Omega=\Omega(t)$ is periodic of period $2\pi/|\omega_0|$, for  $\omega_0 \neq 0 $ and it has the form
    \begin{align} \label{eq:synchronised}
\Omega(t) = \left( \cos(\theta_0 + \omega_0 t), \sin(\theta_0 + \omega_0 t) \right)
    \end{align}
 and $\Omega(t=0,x)=(\cos(\theta_0),\sin(\theta_0))$.
\end{prp}

\begin{cor}[Solutions for constant initial data] \label{cor:constant_initial_data}
    Assume that uniqueness of solutions holds for the system \eqref{eq:macro_eq1}-\eqref{eq:macro_eq3}. If we consider constant initial data $(\rho_0, \Omega_0,\omega_0)$, then the solution has synchronised behaviour with $\rho(t,x)=\rho_0$, $\bar\omega_0(t,x)=\omega_0$ and $\Omega=\Omega(t)$ given in \eqref{eq:synchronised}.
\end{cor}
This statement is direct from the previous Prop. \ref{prop:solutions_constant_density}.

\begin{proof}[Proof of Prop. \ref{prop:solutions_constant_density}]
We plug the identities
\[
\partial_t \Omega = \dot{\theta}(t) \, \Omega^\perp(t),
\qquad
\text{where} \quad \Omega^\perp(t) := (-\sin \theta(t), \cos \theta(t)).
\]
into \eqref{eq:macro_eq3} and obtain
\[
\rho_0 \left(  \partial_t \Omega
-  \omega_0 \, \Omega^\perp \right) = 0\,.
\]
The latter reduces to
\[
\rho_0 \left( \dot{\theta}(t) - \omega_0 \right) \Omega^\perp(t) = 0.
\]
Since \( \rho_0 > 0 \), \( \Omega^\perp(t) \neq 0 \), we conclude
\[
\dot{\theta}(t) = \omega_0.
\]
and integrating in time yields
\[
\theta(t) = \theta_0 + \omega_0 t.
\]
Therefore, the orientation evolves as
\[
\Omega(t) = \left( \cos(\theta_0 + \omega_0 t), \sin(\theta_0 + \omega_0 t) \right).
\]
and the motion is clearly periodic in time with period
$
T = \frac{2\pi}{|\omega_0|}$.

\end{proof}

\begin{remark}
    Also, since the total angular velocity is preserved we have that, in the case of the torus $\mathbb{T}$:
$$\int_{\mathbb{T}}(\rho\omega)(t,x) \ dx= \int_{\mathbb{T}}(\rho\omega)(t=0,x) \ dx=\rho_0\omega_0|\mathbb{T}|,$$
where $|\mathbb{T}|$ is the dimension of the torus. This implies that from the initial data we can predict the final angular velocity $\omega_0$ and the period $T$, assuming that the system converges to this type of periodic solutions. 
\end{remark}

\subsection{Proof of Th. \ref{thm:main_macro}}

\subsubsection{The operator $Q$}

We next summarize key properties of the operator $Q$ in \eqref{eq:collision_operator}, obtained by directly adapting the results of \cite{degond2017continuum}.

\begin{lem}\label{collisional_operator}
$(i)$  $Q$ can be written as 
\begin{equation}\label{eq:collisional_op_2}
    Q(f^\varepsilon) = I_\theta(f^\varepsilon) + I_\omega(f^\varepsilon)\,,
\end{equation}
where
\begin{equation}   
    I_{\omega}(f) = \beta^2 \p_{\omega}\Big[\mathcal{M}_{\overline{\omega}_f}\p_{\omega}\Big(\frac{f}{\mathcal{M}_{\overline{\omega}_f}}\Big)\Big]\,, \qquad 
    I_{\theta}(f) = \alpha^2 \p_{\theta}\Big[\mathcal{N}_{\overline{\theta}_f}\p_{\theta}\Big(\frac{f}{\mathcal{N}_{\overline{\theta}_f}}\Big)\Big]\,.
\end{equation}
\noindent (ii) Define the dissipation functional:
    \begin{equation}
        \mathcal{D}(f) := \int_{(-\pi,\pi] \times \mathbb{R}} Q(f)\, \frac{f}{\mathcal{N}_{\overline{\theta}_f}(\theta)\M_{\overline{w}_f }(\omega)}\, d\theta\, d\omega.
    \end{equation}
    Then we have:
\begin{equation}
\begin{aligned}
\mathcal{D}(f)
&= - \int_{(-\pi,\pi] \times \mathbb{R}}
\mathcal{N}_{\overline{\theta}_f}(\theta)\,
\mathcal{M}_{\overline{w}_f}(\omega) \\
&\quad \times \Bigg[
\alpha^2 \left| \partial_\theta
\left( \frac{f}{
\mathcal{N}_{\overline{\theta}_f}(\theta)\,
\mathcal{M}_{\overline{w}_f}(\omega)} \right) \right|^2
+ \beta^2 \left| \partial_\omega
\left( \frac{f}{
\mathcal{N}_{\overline{\theta}_f}(\theta)\,
\mathcal{M}_{\overline{w}_f}(\omega)} \right) \right|^2
\Bigg] \, d\theta\, d\omega \le 0.
\end{aligned}
\end{equation}
\noindent (iii) The equilibria of \(Q\) (i.e., the functions \(f = f(\theta, \omega) \geq 0\) such that \(Q(f) = 0\)) form a three-dimensional manifold \(\mathcal{E}\) given by:
\[
\mathcal{E} = \left\{ \rho\, \mathcal{N}_{\bar{\theta}}(\theta)\, \mathcal{M}_{\bar{\omega}}(\omega) \;\middle|\; \rho \in \mathbb{R}_+,\, \bar{\theta} \in (-\pi, \pi],\, \bar{\omega} \in \mathbb{R} \right\},
\]
where \(\rho\) is the total mass, \(\bar{\theta}\) is the angle of the mean orientation of motion, and \(\bar{\omega}_f\) is the mean angular velocity of \(\rho\, \mathcal{N}_{\bar{\theta}}\, \mathcal{M}_{\bar{\omega}}\).
\end{lem}

\subsubsection{Generalized Collisional Invariant}
To derive the hydrodynamic limit of \( f^\varepsilon \), we must identify the collisional invariants of the operator \( Q \), i.e., functions \( \psi \) such that
\begin{equation}\label{eq:CI}
\int Q(f)\, \psi\, d\theta\, d\omega = 0, \quad \text{for all } f.
\end{equation}
It is straightforward to verify that \( \psi = 1 \) is a collisional invariant, as it reflects the conservation of total mass. In our setting, the function \( \psi = \omega \) is also a collisional invariant, corresponding to the conservation of total angular momentum. Beyond these, however, the collision operator admits no other standard invariants.
Yet, since the equilibrium manifold of \( Q \) is three-dimensional, we need an additional conserved quantity to derive a closed macroscopic system. To overcome this difficulty, we use the notion of \emph{Generalized Collisional Invariant (GCI)} introduced in \cite{degondcontinuum}. The key idea is to relax condition \eqref{eq:CI}.
\begin{dfn}
We define $\mathcal{I}(f;\overline{\theta})$ for a given $\bar \theta \in (-\pi, \pi]$ by:
\[
\mathcal{I}(f;\overline{\theta}) = \partial_\theta \left( k_\theta \sin(\theta - \bar{\theta}) f + \alpha^2 \partial_\theta f \right).
\]
\end{dfn}

\begin{dfn}[Generalized Collisional Invariant \cite{degondcontinuum}]
Let \( \bar{\theta} \in (-\pi, \pi] \) be fixed and define \( \Omega = (\cos \bar{\theta}, \sin \bar{\theta}) \). We say that a function \( \chi_{\bar{\theta}}: (-\pi, \pi] \to \mathbb{R} \) is a \emph{Generalized Collisional Invariant (GCI)} associated to \( I_\theta \) at orientation \( \Omega \), if and only if:
\begin{equation}\label{def:GCI}
\int_{(-\pi,\pi] } \mathcal{I}(f;\bar{\theta})\, \chi_{\bar{\theta}}(\theta)\, d\theta = 0 \quad \mbox{ for all } f \,\text{such that}\quad 
\int_{(-\pi,\pi]} f(t,x,\theta,\omega) \, \sin(\theta - \bar{\theta}) \, d\theta\, = 0.   
\end{equation}
\end{dfn}

Notice that the condition of $f$ in \eqref{def:GCI} is equivalent to    
\begin{equation}
        \Vec{\tau}(\bar\theta)^\perp\cdot\int_{(-\pi,\pi]} \Vec{\tau}(\theta) f(t,x,\theta,\omega)\,  d\theta = 0.
    \end{equation}
For the particular case in which $\bar \theta = \bar \theta_{f^\varepsilon}$ and $f= f^\varepsilon$, the condition above holds, i.e.
\begin{equation}
        \Vec{\tau}(\bar\theta_{f^\varepsilon})^\perp\cdot\int_{(-\pi,\pi]} \Vec{\tau}(\theta) f^\varepsilon(t,x,\theta,\omega)\,  d\theta = 0,
    \end{equation}
and, therefore, by the definition of the Generalized Collision Invariant we have that
\begin{equation}\label{eq:GCI_condition}
  0=\int_{(-\pi,\pi] } \mathcal{I}(f^\varepsilon;\bar{\theta}_{f^\varepsilon})\, \chi_{\bar{\theta}_{f^\varepsilon}}(\theta)\, d\theta=\int_{(-\pi,\pi] } I_\theta (f^\varepsilon)  \chi_{\bar{\theta}_{f^\varepsilon}} \, d\theta \, \qquad \forall \varepsilon>0.  
\end{equation} 
This last equality is key to computing the mean orientation in the limit. What is left is to characterise the Generalized Collision Invariant.

\medskip

\begin{prp}[From \cite{degondcontinuum,frouvelle2012continuum}]
Let $\Omega \in \mathbb{S}^1$ be a fixed orientation and write $\Omega = (\cos\bar\theta, \sin\bar\theta)$ for some $\bar\theta \in (-\pi, \pi]$. Then the space of Generalized Collisional Invariants (GCI) associated to $\Omega$ is the two-dimensional vector space
\[
C_\Omega = \mathrm{Span}\{1, \chi_{\overline{\theta}}\},
\]
where $\chi_{\overline{\theta}} = g(\theta - \bar\theta)$ and $g$ is an odd function defined as 
\begin{equation}
    g(\Gamma) = \frac{\alpha^2}{k_\theta}\Gamma - \frac{\alpha^2}{k_\theta}\pi\,\frac{\int_0^\Gamma \, \exp(\frac{-k_\theta\cos(\varphi)}{\alpha^2})d\varphi}{\int_0^\pi \, \exp(\frac{-k_\theta\cos(\varphi)}{\alpha^2})d\varphi}.
\end{equation}

\end{prp}
We will use $\chi_{\overline{\theta}}$ to close the macroscopic equation for the orientation field $\Omega(t,x)$ by integrating the kinetic equation against $\chi_\Omega(\theta)$ and passing to the limit $\varepsilon \to 0$.

\subsubsection{Proof of theorem \ref{thm:main_macro}}\label{sec:limit}
\begin{remark}
     When the meaning is clear from the context, we will omit writing explicitly the integration domain \( (-\pi, \pi] \times \mathbb{R} \) in integrals over \( (\theta, \omega) \).
\end{remark}
By assumption, \( f^\varepsilon \) converges to \( f^0 \) as \( \varepsilon \to 0 \). Taking the limit in the rescaled kinetic equation \eqref{eq:kinetic_scaled1} we have that
\[
Q(f^0) = 0,
\]
which means that \( f^0 \) is in the kernel of $Q$. By Lemma \ref{collisional_operator}, \( f^0 \) is of the form
\[
f^0 = \rho^0 \, \mathcal{N}_{\bar{\theta}_0}(\theta) \, \mathcal{M}_{\bar{\omega}_0}(\omega),
\]
where $\rho^0 = \rho^0(t,x)$, $\theta_0 = \theta_0(t,x)$, and $\bar\omega_0 = \bar \omega_0(t,x)$. Moreover \( \mathcal{N}_{\bar{\theta}_0} \) and \( \mathcal{M}_{\bar{\omega}_0} \) are the normalized distributions defined in \eqref{eq:equilibria}.

\medskip
\paragraph{Equation for the density $\rho^0$.}
To determine the evolution of the unknown \( \rho^0 \), we first integrate the kinetic equation over \( (\theta, \omega) \) and obtain the conservation equation:
\[
\partial_t \rho^\varepsilon + \nabla_x \cdot {j}^\varepsilon = 0,
\]
where the flux \( {j}^\varepsilon \) is defined by:
\[
{j}^\varepsilon = \int \vec{\tau}(\theta)\, f^\varepsilon \, d\theta \, d\omega.
\]
Because \(\int_{\mathbb R}\mathcal M_{\bar\omega_0}(\omega)\,d\omega=1\),
in the limit for $\varepsilon \to 0$, we obtain
\[
j^{0}
   \;=\;
   \rho^{0}\!
   \int_{-\pi}^{\pi}
        \vec\tau(\theta)\,
        \mathcal N_{\bar\theta_0}(\theta)\,d\theta
   \;=\;
   \rho^{0}\!
   \int_{-\pi}^{\pi}
        \vec\tau(\theta)\,
        \frac{e^{\kappa\cos(\theta-\bar\theta_0)}}{2\pi I_{0}(\kappa)}
        \,d\theta,
\]
where $\kappa = k_\theta/\alpha^2$ and
\begin{align} \label{eq:integralc1}
\int_{-\pi}^{\pi}
        \vec\tau(\theta)\,
        \frac{e^{\kappa\cos(\theta-\bar\theta_0)}}{2\pi I_{0}(\kappa)}
        \,d\theta = c_1 \Omega^0\, \quad \mbox{ for } \quad c_1:= \frac{I_{1}(\kappa)}{I_{0}(\kappa)}, 
\end{align}
where $\Omega^0= (\cos \bar\theta_0, \sin \bar\theta_0).$ This equality is proven in Appendix \ref{sec:proof_integral}.

Hence, in the limit \( \varepsilon \to 0 \), we obtain:
\[
{j}^\varepsilon \xrightarrow{\varepsilon \to 0} {j}^0 = c_1 \rho^0 \Omega^0\,.
\]
Therefore, \( \rho^0 \) satisfies the macroscopic mass conservation equation
\[
\partial_t \rho^0 + c_1 \nabla_x \cdot (\rho^0 \Omega^0) = 0.
\]

\medskip
\paragraph{Equation for the mean angular velocity $\bar\omega^0$.}
To derive the evolution equation for the mean angular velocity $\overline{\omega}^0$, we multiply the kinetic equation \eqref{eq:kinetic_scaled1} by $\omega$ and integrate over $(\theta, \omega)$.  We obtain 
\[
\partial_t (\rho^\varepsilon\overline{\omega}^\varepsilon) + \nabla_x \cdot \left( \int \omega\, \vec{\tau}(\theta)\, f^\varepsilon\, d\theta\, d\omega \right) = \frac{1}{\varepsilon} \int \omega\, Q(f^\varepsilon)\, d\theta\, d\omega + \mathcal{O}(\varepsilon).
\]
The right-hand side integral is equal to zero for all $\varepsilon$, indeed 
\begin{equation}
\begin{aligned}
\int \omega\,Q(f^\varepsilon)\,d\theta\,d\omega
  =\, &k_\theta \int \omega\,\partial_\theta\!\bigl(f^\varepsilon F_{f^\varepsilon}\bigr)\,d\theta\,d\omega
    + \alpha^{2}\int \omega\,\partial_{\theta}^{2}f^\varepsilon\,d\theta\,d\omega\\
    + &k_\omega \int \omega\,\partial_\omega\!\bigl(f^\varepsilon G_{f^\varepsilon}\bigr)\,d\theta\,d\omega
    + \beta^{2}\int \omega\,\partial_{\omega}^{2}f^\varepsilon\,d\theta\,d\omega \,,   
\end{aligned}
\end{equation}
where the first two terms are zero by the fact that $f$ is $2\pi-$periodic and the fundamental theorem of calculus. For the third term we integrate by part 
and since \(G_f=\bar\omega_f-\omega\) we have
\begin{equation*}
    \int \omega\,\partial_\omega\!\bigl(f^\varepsilon G_{f^\varepsilon}\bigr)\,d\theta\,d\omega= - \int (\bar\omega_f-\omega)\,f^\varepsilon \d\theta \d\omega = -\, \bar \omega_f \rho^\varepsilon - \int \omega f^\varepsilon d\theta d\omega =0.
\end{equation*}
Similarly to the first two terms, also the term multiplying $\beta^2$ is equal to zero.
Moreover, in the limit for $\varepsilon \to 0$, we have that
\[
\int \omega\, \vec{\tau}(\theta)\, f^\varepsilon\, d\theta\, d\omega \ \xrightarrow{\varepsilon \to 0}  \rho^0 \overline{\omega}^0 \int \vec{\tau}(\theta)\, \mathcal{N}_{\bar{\theta}_0}(\theta)\, d\theta = c_1 \rho^0 \overline{\omega}^0 \Omega^0,
\]
since $\overline{\omega}^0= \int \omega\mathcal{M}_{\overline{\omega}_0}$ and $c_1$ is defined as before.
We thus obtain the second macroscopic equation:\label{eq:macro_eq2_dim}
\begin{equation}
\partial_t (\rho^0 \overline{\omega}^0) + c_1 \nabla_x \cdot (\rho^0 \overline{\omega}^0 \Omega^0) = 0.   
\end{equation}

\medskip
\paragraph{Equation for the mean orientation $\Omega^0$.}
Now to obtain the equation for the mean orientation $\Omega^0 = \Omega^0(t,x)$, we multiply \eqref{eq:kinetic_scaled1} by the GCI $\chi_{\overline{\theta}_\varepsilon}$, where $\Omega^\varepsilon = \Omega_{f^\varepsilon}$, and integrate with respect to $\theta$ and $\omega$. By \eqref{eq:GCI_condition} we have that
\begin{equation}
    \int Q(f^\varepsilon) \chi_{\overline{\theta}_\varepsilon} \d\theta \, \d\omega = 0\,.
\end{equation}
Indeed, this holds since
\begin{equation}
    \int I_\theta(f^\varepsilon)\chi_{\overline{\theta}_\varepsilon}\,\d\theta = \int \mathcal{I}(f^\varepsilon; \overline{\theta}_f^\varepsilon)\,\chi_{\overline{\theta}_\varepsilon}\d\theta = 0\,,
\end{equation}
given that $f^\varepsilon$ satisfies the condition in \eqref{def:GCI} for $\bar \theta = \bar \theta^\varepsilon_f$.

Hence we obtain that 
\begin{equation}
    \int \Big( \p_t f^\varepsilon + \Vec{\tau}(\theta) \cdot \nabla_x f^\varepsilon +  \omega \p_{\theta}f^\varepsilon\Big)\,   \chi_{\overline{\theta}\varepsilon}(\theta) d\theta \,d\omega= 0\,,
\end{equation}
which, in a more compact form, can be written as 
\begin{equation}
    \int \Big(\mathcal{T}^1 f^\varepsilon + \mathcal{T}^2 f^\varepsilon + \mathcal{T}^3 f^\varepsilon \Big)\chi_{\overline{\theta}_\varepsilon}(\theta) \, d\theta\,d\omega = 0\,,
\end{equation}
where $\mathcal{T}^1, \mathcal{T}^2$, and $\mathcal{T}^3$ are defined by the following operators 
\begin{equation}
\begin{aligned}
    &\mathcal{T}^1 \, f= \p_t f\,,\\
    & \mathcal{T}^2 \, f= \Vec{\tau}(\theta) \cdot \nabla_x f\,,\\
    & \mathcal{T}^3 \, f=\omega \p_{\theta}f\,.\\
    \end{aligned} 
\end{equation}
We note that
$\chi_{\overline{\theta}_\varepsilon}$ is smooth enough
and therefore $\chi_{\overline{\theta}_\varepsilon}\to \chi_{\overline{\theta}_0}$ for $\varepsilon \to 0$. Hence in the limit $\varepsilon\to 0$ we get:
\begin{equation} \label{eq:limit_GCI_explicit}
\int
\big[
\partial_t \left( \rho\, \mathcal{N}_{\bar{\theta}_0}(\theta)\, \mathcal{M}_{\bar{\omega}_0}(\omega) \right)
+ \vec{\tau}(\theta) \cdot \nabla_x \left( \rho\, \mathcal{N}_{\bar{\theta}_0}(\theta)\, \mathcal{M}_{\bar{\omega}_0}(\omega) \right)
+  \omega \partial_\theta \left(\, \rho\, \mathcal{N}_{\bar{\theta}_0}(\theta)\, \mathcal{M}_{\bar{\omega}_0}(\omega) \right)
\big]
\chi_{\overline{\theta}_0}\, d\theta\, d\omega = 0.
\end{equation}
We start to compute the contribution of the operator $\mathcal{T}^1$ and separate each term, obtaining
\[
\mathcal{T}^1(\rho\, \mathcal{N}_{\bar{\theta}_0}(\theta)\, \mathcal{M}_{\bar{\omega}_0}(\omega)) = (\partial_t \rho^0) \mathcal{N}_{\bar{\theta}_0} \mathcal{M}_{\bar{\omega}_0} 
+ \rho^0\, (\partial_t \mathcal{N}_{\bar{\theta}_0}) \mathcal{M}_{\bar{\omega}_0} 
+ \rho^0\, \mathcal{N}_{\bar{\theta}_0} (\partial_t \mathcal{M}_{\bar{\omega}_0})\,,
\]
where the first term vanishes upon integration against \( \chi_{\bar{\theta}} \), due to the odd symmetry of \( g(\theta - \bar{\theta}) \) and the even symmetry of \( \mathcal{N}_{\bar{\theta}} \). The third term vanishes due to the identity
\[
\int (\omega - \bar{\omega}) \mathcal{M}_{\bar{\omega}}(\omega)\, \d\omega = 0.
\]
Therefore, the only non zero term is 
\begin{align}
\int \mathcal{T}^1(\rho\, \mathcal{N}_{\bar{\theta}_0}(\theta)\, \mathcal{M}_{\bar{\omega}_0}(\omega)) \chi_{\bar\theta_0}\,\d\theta\d\omega
&= \frac{k_\theta}{\alpha^2}\rho^0\, \partial_t \bar{\theta}_0 \int \sin(\theta - \bar{\theta})\, \mathcal{N}_{\bar{\theta}}(\theta)\, g(\theta - \bar{\theta})\, \d\theta \nonumber\\
& =\frac{k_\theta}{\alpha^2}\,\rho^0\, \p_t \overline{\theta}_0 \int \sin(\theta)\, \N_{0} \,g(\theta)\, \d\theta \nonumber\\
& =K_1\,\frac{k_\theta}{\alpha^2}\,\rho^0\, \p_t \overline{\theta}_0\,, \label{eq:T1}
\end{align}
where in the last equality we have performed a change of variables, used the fact that $g$ is a $2\pi-$periodic function 
and $K_1$ defined in \eqref{def:K1}.
We now compute the transport term
\begin{align*}
\mathcal{T}^2(\rho\, \mathcal{N}_{\bar{\theta}_0}(\theta)\, \mathcal{M}_{\bar{\omega}_0}(\omega)) 
= &\big(\vec{\tau}(\theta) \cdot \nabla_x \rho^0\big)\mathcal{N}_{\bar{\theta}_0}(\theta)\mathcal{M}_{\bar{\omega}_0}(\omega)
+ \rho^0\, (\vec{\tau}(\theta) \cdot \nabla_x \bar{\theta}_0)\, \partial_{\bar{\theta}} \mathcal{N}_{\bar{\theta}_0}(\theta)\, \mathcal{M}_{\bar{\omega}_0}(\omega)\\
&+ \rho^0\, \mathcal{N}_{\bar{\theta}_0}(\theta)\, (\vec{\tau}(\theta) \cdot \nabla_x \bar{\omega}_0)\, \partial_{\bar{\omega}} \mathcal{M}_{\bar{\omega}_0}(\omega)=:\mathcal{T}^2_{1}+\mathcal{T}^2_{2}+\mathcal{T}^2_{3}.
\end{align*}
We use the decomposition
$
\vec{\tau}(\theta) = \cos(\Gamma)\, \Omega^0 + \sin(\Gamma)\, (\Omega^0)^\perp
$, where $\Gamma=(\theta-\overline{\theta})$  and $(\Omega^0)^\perp = (-\sin \theta_0, \cos \theta_0)$, which leads to
\begin{equation}
\begin{aligned}
\int \mathcal{T}^2_{1} \chi_{\bar\theta_0}\d\theta\d\omega &=
\int(\vec{\tau}(\theta) \cdot \nabla_x \rho)\, \mathcal{N}_{\bar{\theta_0}}\, \mathcal{M}_{\bar{\omega_0}}\, g(\theta -\overline{\theta})\, d\theta\, d\omega \\
&= \int(\vec{\tau}(\theta) \cdot \nabla_x \rho)\, \mathcal{N}_{\bar{\theta}_0}\, g(\theta -\overline{\theta})\, d\theta\\
&= \nabla_x  \rho^0 \cdot \int\sin(\theta -\overline{\theta})\N_{\overline{\theta}_{0}}g(\theta -\overline{\theta})\,d\theta \, ({\Omega}^0)^\perp\,,\\
&=K_1 \nabla_x \rho^0 \cdot(\overline{\Omega}^0)^\perp\,, 
\end{aligned}
\end{equation}
due to the symmetry of the integrand.
The second term gives:
\begin{equation}
\begin{aligned}
\int \mathcal{T}^2_{2}\chi_{\bar\theta_0} \d\theta\d\omega &= \int\rho^0\, (\vec{\tau}(\theta) \cdot \nabla_x \bar{\theta}_0)\, \partial_{\bar{\theta}} \mathcal{N}_{\bar{\theta}_0}(\theta)\, \mathcal{M}_{\bar{\omega}_0}(\omega)\, g(\theta -\overline{\theta})\, d\theta\, d\omega\\
&= \rho^0\frac{k_\theta}{\alpha^2} \nabla_x  \overline{\theta}_{0} \cdot  \int\vec{\tau}(\theta)\, \sin(\theta - \overline{\theta})\mathcal{N}_{\bar{\theta}_0}(\theta)\, g(\theta -\overline{\theta})\, d\theta\\
&=\rho^0\frac{k_\theta}{\alpha^2} \nabla_x  \overline{\theta}_{0} \cdot \int(\cos\Gamma{\Omega}^0 + \sin \Gamma ({\Omega}^0)^\perp) \sin\Gamma\, \N_{\overline{\theta}_{0}} g(\Gamma) d\theta\\
&= \rho^0\frac{k_\theta}{\alpha^2} \nabla_x \overline{\theta}_{0} \cdot \int\cos(\theta -\overline{\theta})\, \sin(\theta -\overline{\theta})\N_{\overline{\theta}_{0}} g(\theta -\overline{\theta}) \, d\theta \, {\Omega}^0\\
&= \rho^0\, \frac{k_\theta}{\alpha^2} K_2\, (\nabla_x \bar{\theta}_0 \cdot \Omega^0),
\end{aligned}
\end{equation}
with $K_2$ defined in \eqref{def:K1}.
Finally, $\int \mathcal{T}^2_{3}\chi_{\bar\theta_0}\d\theta\d\omega = 0$
since $
\int_{\omega} \partial_{\bar{\omega}} \mathcal{M}_{\bar{\omega}_0}(\omega)\, d\omega = 0.
$
Therefore, the total contribution from the transport term is
\begin{align} \label{eq:T2}
\int \mathcal{T}^2 (\rho\, \mathcal{N}_{\bar{\theta}_0}(\theta)\, \mathcal{M}_{\bar{\omega}_0}(\omega)) \chi_{\bar\theta_0}\d\theta\d\omega
= K_1\, \nabla_x \rho^0 \cdot (\Omega^0)^\perp
+ \rho^0\, \frac{k_\theta}{\alpha^2} K_2\, (\nabla_x \bar{\theta}_0 \cdot \Omega^0)\,.
\end{align}

We now compute the contribution of the angular transport term $\mathcal{T}_3$:
\begin{align}
\int \mathcal{T}_3(\rho\, \mathcal{N}_{\bar{\theta}_0}(\theta)\, \mathcal{M}_{\bar{\omega}_0}(\omega)) \chi_{\bar\theta_0}\d\theta\d\omega &=\int \omega \,\mathcal{M}_{\bar{\omega}_0} \,\partial_\theta \left(\, \rho\, \mathcal{N}_{\bar{\theta}_0}(\theta)\, (\omega) \right)
\big) g(\theta - \bar{\theta}_0)\, d\theta\, d\omega\nonumber\\
&=- \frac{k_\theta}{\alpha^2} \bar\omega^0 \rho^0 \int \sin(\theta- \overline{\theta}_{0}) \N_{\overline{\theta}_{0}}g(\theta-\overline{\theta}_{0})\, d\theta \nonumber\\
&=- K_1\frac{k_\theta}{\alpha^2} \bar\omega^0 \rho^0. \label{eq:T3}
\end{align}

Finally, putting together \eqref{eq:T1}, \eqref{eq:T2} and \eqref{eq:T3} yields:
\begin{equation}\label{eq:final_result}
K_1\,\frac{k_\theta}{\alpha^2}\,\rho^0\, \p_t \overline{\theta}_0 +K_1 \nabla_x \rho^0 \cdot(\Omega^0)^\perp +\rho^0\, \frac{k_\theta}{\alpha^2} K_2\, (\nabla_x \bar{\theta}_0 \cdot \Omega^0)- K_1\frac{k_\theta}{\alpha^2} \bar\omega^0 \rho^0 =0.
\end{equation}
Using that vector \( \Omega^0 = \vec{\tau}(\bar{\theta}_0) \), elementary computations show that:
\begin{equation}\label{eq:equality_Omega}
\partial_t \Omega^0 = \partial_t \bar{\theta}_0 \, (\Omega^0)^\perp 
\qquad \text{and} \qquad
(\Omega^0 \cdot \nabla_x) \Omega^0 = ((\Omega^0)^\perp \otimes \Omega^0)\, \nabla_x \bar{\theta}_0. 
\end{equation}
Therefore, multiplying equation~\eqref{eq:final_result} by \( (\Omega^0)^\perp \) leads finally to (dropping the superscript for simplicity of notation):
\begin{equation}\label{eq:macro_orientation}
    \rho\Big( K_1 \frac{k_\theta}{\alpha^2 }\p_t \Omega + K_2\frac{k_\theta}{\alpha^2 } (\Omega \cdot \nabla_x) \Omega - K_1\frac{k_\theta}{\alpha^2 }\overline{\omega}\, \Omega ^\perp\Big) + K_1P_{\Omega^\perp}\nabla_x \rho = 0\,,
\end{equation}
where $P_{\Omega^\perp} = \mathrm{Id} - \Omega \otimes \Omega$ is the projection operator onto $\Omega^\perp$. Dividing by $K_1 k_\theta/\alpha^2$ we obtain equation \eqref{eq:macro_eq3} for $\Omega$.
This concludes the formal derivation of the macroscopic system stated in Theorem~\ref{thm:main_macro}.

\section{Simulations}
\subsection{Simulation of the microscopic VK model}\label{sec:simulation_micro}

\subsubsection{Numerical integration of the IBM}
In this section, we provide additional details on the numerical methods used to produce the simulations on the IBM \eqref{eq:VicsekKuramoto1}–\eqref{eq:VicsekKuramoto3} shown in section \ref{sec:phase_diagrams}. The code is freely available on the GitHub repository of the second author~\cite{carmela_moschella_2025_17945342}.

We simulate the IBM \eqref{eq:VicsekKuramoto1}–\eqref{eq:VicsekKuramoto3} in dimension \( n = 2 \) on a periodic square domain. The time integration is performed using a modified Euler–Maruyama scheme with adaptive time stepping. The adaptive time stepping is significant in this setting due to the potential stiffness introduced by certain parameter regimes. The implementation uses the \texttt{StochasticDiffEq.jl} package in Julia~\cite{rackauckas2017differentialequations}.

Although simulating mean-field particle systems is conceptually straightforward, the computational cost becomes significant when the number of particles is large, especially due to the computation of the interaction terms. To address this issue, we implement a Verlet neighbor list method that restricts the computation of the interaction forces to particles located within a fixed cutoff radius. The spatial domain is partitioned into a regular grid of cells, and during each update, interaction forces are computed only among particles that belong to the same or adjacent cells. This approach avoids unnecessary distance checks and reduces the computational complexity from $\mathcal{O}(N^2)$ to approximately $\mathcal{O}(N)$. The method is implemented using the \texttt{CellListMap.jl} package~\cite{martinez2022celllistmap}, which provides optimized routines for evaluating interaction kernels in  two and three-dimensional periodic domains.

Simulations are conducted in a periodic square \( [0, L_x] \times [0, L_y] \) with \( L = L_x = L_y = 64 \), and the number of particles is fixed at \( N = 15,\!000 \). The initial positions $(x_1, x_2)_i$ and directions $\theta_i$ of the particles
are taken uniform and random for $i = 1,..., N$ in $[0,L]$ and $[0,2\pi] $ respectively, while the initial angular velocities \( \omega_i \) are sampled from an asymmetric distribution. Indeed for a quarter of the particles we prescribe \( \omega_i = 5 \cdot \mathrm{rand}() \), where \texttt{rand()} denotes a uniform random variable on \([0,1]\). For the remaining three-quarters, we set \( \omega_i = -5 \cdot \mathrm{rand}() \). This choice ensures that the local mean angular velocity is different from zero, thereby avoiding the particles relaxing toward a non-rotating state, which would prevent the formation of the patterns discussed in Section \ref{sec:phase_diagrams}. All the simulations are run with a time step \( \Delta t = 0.01 \) and integrated up to a final time \( t_{\text{end}} = 0.01 \times 2\cdot 10^4 \). Please refer to table \ref{tab:parameters_micro} for a full list of parameter values.

\begin{table}[ht]
\centering
\caption{Parameters used for the microscopic simulations. The values considered for \(k_\theta\) and \(k_\omega\) are specified in the caption of the corresponding figures.}
\begin{tabular}{lll}
\toprule
\textbf{parameters} & \textbf{value} & \textbf{description} \\
\midrule
\(N\)       & 15\,000     & total number of individuals \\
\(k_\theta\) & various     & interaction strength in the direction variable \\
\(k_\omega\) & various     & interaction strength in the angular velocity variable \\
\(t_{\text{end}}\) & 200     & final time of the simulation \\
\(\sqrt{2\alpha^2}\) & 0.5       & angular noise in orientation dynamics \\
\(\sqrt{2\beta^2}\) & 0.5       & noise in angular velocity dynamics \\
\(\Delta t\)      & 0.01        & time step size \\
\(R\)       & 2        & interaction radius \\
\(v_0\)     & 1           & constant velocity of particles \\
\(L\)       & 64          & size of the periodic domain \\
\bottomrule
\end{tabular}
\label{tab:parameters_micro}
\end{table}

\subsubsection{Patterns: phase diagrams}\label{sec:phase_diagrams}
\begin{figure}[h!]
\centering
\includegraphics[width=12cm, height =12cm]{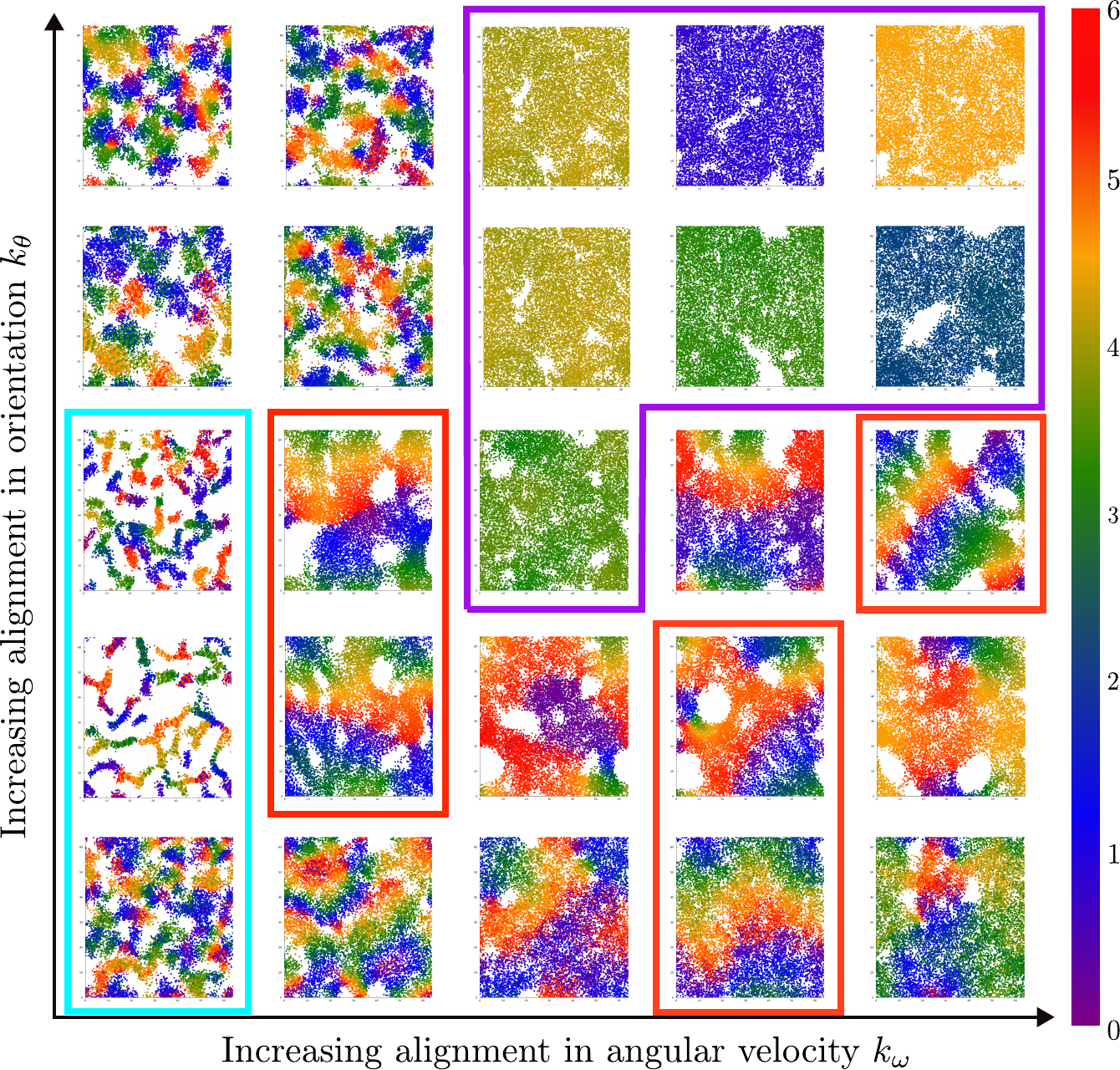}
\caption{Results of microscopic simulations for varying values of the alignment parameters \(k_{\theta}\) and \(k_{\omega}\), with all other parameters fixed as reported in Table \ref{tab:parameters_micro}. Each simulation involves \(15{,}000\) agents interacting within an interaction radius \(R = 2\). The columns (from left to right) correspond to increasing values of the angular velocity alignment strength \(k_{\omega} = 1, 11, 51, 61, 81\), while the rows (from bottom to top) correspond to increasing values of the directional alignment strength \(k_{\theta} = 1, 11, 21, 61, 71\). The colormap, shown on the right hand side of the figure, provides the correspondence between orientation angles and colors. We can distinguish at least three types of patterns: rotating clusters (framed in light blue), traveling waves in orientation (framed in red), synchronised behaviour (framed in purple). Videos are available in the appendix.}
\label{fig:phase_diagram}
\end{figure}

In this section, we present the results of the numerical simulations of the IBM~\eqref{eq:VicsekKuramoto1}–\eqref{eq:VicsekKuramoto3} for various values of the alignment parameters $k_\theta$ and $k_\omega$. In particular, 
Figure~\ref{fig:phase_diagram} shows the output of the simulations after time $t^{*} =0.01\times1.9\cdot 10^4$. At time $t^{*}$, the system
has reached a steady state. In these figures, particles are represented as dots, and their color gives their orientation.
The simulation results are organized in a table format, with columns (from left to right) corresponding to increasing values of the angular velocity alignment strength \( k_\omega \), and rows (from bottom to top) corresponding to increasing values of the orientation alignment coefficient \( k_\theta \). Moreover we refer the reader to the caption of Figure~\ref{fig:phase_diagram} for the precise values of \( k_\theta \) and \( k_\omega \) used in each simulation. Among all the results, we select three representative cases that illustrate the main qualitative patterns observed across the parameter space. Videos corresponding to these cases are provided in the supplementary material (Appendix~\ref{appendix:videos}), where additional details are provided.

\begin{figure}[h!]
\centering
\includegraphics[width=17.2cm, height=5cm]{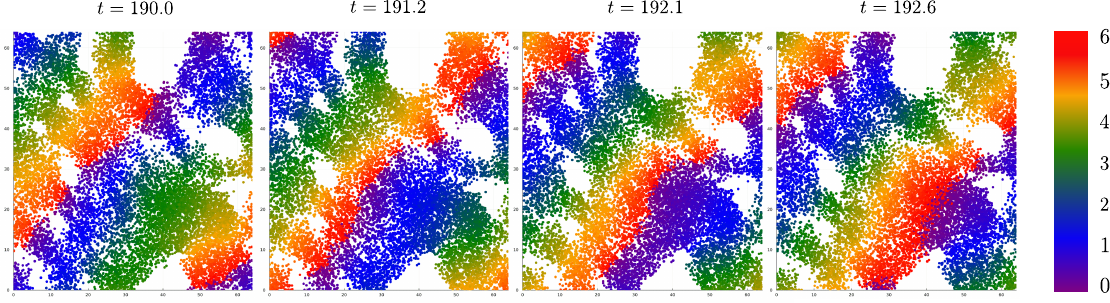}
\caption{
Sequence of snapshots showing the emergence of a traveling wave in orientation for alignment parameters \( k_\theta = 21 \) and \( k_\omega = 81 \). The system evolves towards a state with nearly uniform spatial density, while the orientation field propagates with a wave-like motion across the domain.
The corresponding video can be found in Appendix \ref{appendix:videos} (Video 2).}
\label{fig:traveling_micro}
\end{figure}

\begin{figure}[h!]
\centering
\includegraphics[width=17.2cm, height=5cm]{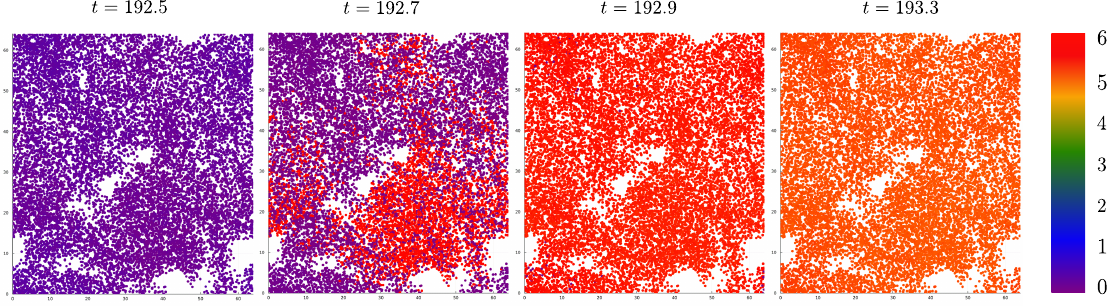} 
\caption{
Sequence of snapshots illustrating the steady-state dynamics for alignment parameters \( k_\theta = k_\omega = 71 \). To highlight the synchronised behavior, we display multiple time frames from the same simulation. The particles rotate collectively as a rigid body, with the orientation field evolving in time across the entire domain. This global rotation characterizes the synchronised regime, where the direction of motion changes uniformly for all the particles in the domain, while the distribution of particles remains uniform in space. The corresponding video can be found in Appendix \ref{appendix:videos} (Video 3).
}
\label{fig:synchronised}
\end{figure}

In Figure~\ref{fig:phase_diagram}, we observe a rich spectrum of emergent collective behaviors as we explore the influence of different values of \( k_\theta \) and \( k_\omega \). We classify the observed patterns into three main types, each highlighted in the figure using different colored frames corresponding to distinct regions in parameter space:

\begin{itemize}
    \item \textbf{Rotating clusters} (framed in light blue): Agents self-organize into compact, rotating clusters within the domain. This pattern predominantly arises from weak angular velocity alignment (\( k_\omega = 1 \)) combined with weak to moderate orientation alignment (\( k_\theta = 1, 11, 21 \)).
    
    \item \textbf{Traveling waves in orientation} (framed in red): For small to moderate orientation alignment (\( k_\theta = 1, 11, 21 \)) and moderate to strong angular velocity alignment (\( k_\omega = 11, 61, 81 \)), the agents arrange themselves in configurations where the spatial density remains nearly uniform, but the orientation propagates as a traveling wave in the domain. Snapshots illustrating the traveling wave are shown in Figure~\ref{fig:traveling_micro}.
    
    \item \textbf{Synchronised behavior} (framed in purple): We refer to the synchronised regime as a global, synchronized state characterized by large-scale, time-periodic oscillations in the mean orientation of the system. This regime emerges for large values of both the alignment parameters, typically \( k_\theta = 61, 71 \) and \( k_\omega = 51, 61, 81 \). It represents the most stable and recurrent pattern observed in the regime characterized by strong alignment interactions. Although we show only a representative subset in Figure~\ref{fig:phase_diagram} for clarity of presentation, extensive simulations indicate that for \( k_\theta, k_\omega \gtrsim 50 \), the synchronised regime dominates the long-time dynamics of the system. Snapshots illustrating the synchronised behaviour are shown in Figure~\ref{fig:synchronised}.
\end{itemize}
Finally, the simulations not enclosed in colored frames do not exhibit any clearly ordered or persistent collective behavior.

\begin{remark}[Characterisation of synchronised behaviour in the IBM]
    Assume that $a=b=0$ in \eqref{eq:VicsekKuramoto2}-\eqref{eq:VicsekKuramoto3} (i.e., there is no noise). Suppose that initially the angular velocity of all the particles is equal to some constant value $\omega_i(0)= \omega_0\neq 0$, and all the orientations are initially perfectly aligned, so that $\theta_i(0)=\theta_0$, for all $i=1,\hdots, N$.  In this set up, $\theta_i(t)=\theta_0+\omega_0 t$ and therefore the orientation of motion for all the particles $\tau(\theta_i)$ is time-periodic with period $2\pi/|\omega_0|$.
    This is the discrete version of the macroscopic solutions characterised in Prop. \ref{prop:solutions_constant_density} and is the pattern that what we observe approximately in Fig. \ref{fig:phase_diagram} framed in purple.
\end{remark}

\subsubsection{Simulation of the IBM in the macroscopic limit regime}
\label{sec:IBM_parameter}

To compare qualitatively the macroscopic Vicsek--Kuramoto model~\eqref{eq:macro_eq1}--\eqref{eq:macro_eq3} with the simulations of the IBM~\eqref{eq:VicsekKuramoto_dimensionless1}--\eqref{eq:VicsekKuramoto_dimensionless3}, we must chose parameters according to the scaling regime in Section~\ref{sec:scaling}, which defines the hydrodynamic limit. For the IBM, we fix the time step to \(\Delta t = 0.01\) and the particle speed to \(c = 0.1\), so that particles move in a relatively big domain of side length \(L = 1\).

Assuming an initial uniform particle distribution, the average number of interacting neighbors per particle is given by
\[
N_{\text{neigh}} = \frac{N \pi R^2}{L^2},
\]
where \(R\) denotes the interaction radius. To ensure that the system operates in the mean-field regime, a good approximation is given by \(N_{\text{neigh}} \sim 10^2\). For this reason, we take \(N = 2 \cdot 10^4\) and set \(R = 0.04\).
This choice corresponds to an interaction area of about \(0.5\%\) of the total domain, which guarantees the spatial locality required for the validity of the hydrodynamic limit. As detailed in Section~\ref{sec:scaling}, the interaction radius \(R\) acts as the fundamental scaling parameter from which all other parameters in the particle model are derived. For this reason we scale the simulation parameters as 
\[
k_\theta = \frac{k_\theta'}{R}, \quad 
k_\omega = \frac{k_\omega'}{R}, \quad 
\alpha^2 = \frac{(\alpha')^2}{R}, \quad 
\beta^2 = \frac{(\beta')^2}{R},
\]
with \( k_\theta' \), \( k_\omega' \), \( \alpha' \), and \( \beta' \) dimensionless constants. This scaling ensures that alignment and stochastic effects remain significant despite the shrinking spatial range of interactions.
As dimensionless constants, we choose \( k_\theta' = 1 \), \( k_\omega' = 1 \), and noise intensities \( (\alpha')^2 = 0.125 \), \( (\beta')^2 = 0.125 \). 

\begin{remark} \label{rem:kalpha}
Note that, given the values considered here, we chose \( {k_\theta}/{\alpha^2} = 8 \) in the macroscopic simulations.
\end{remark}

\paragraph{Result:}
For this setup, at the final time \( t_{\text{end}} = 100 \), the system exhibits the same synchronised behavior observed in earlier simulations: the orientation field undergoes uniform rotation, while the spatial density remains approximately constant. A video of the simulation is provided in Appendix~\ref{appendix:videos} (Video~4). This experiment suggests that the macroscopic simulations should, at a minimum, reproduce this pattern.

\subsection{Simulation of the macroscopic VK model}\label{sec:simulation_macro}

\subsubsection{Numerical scheme}
In this section, we present the numerical scheme used to simulate the macroscopic system \eqref{eq:macro_eq1}--\eqref{eq:macro_eq3}. 
Following the methodology introduced in \cite{motsch2011numerical} for the SOH model, it is possible to show that the macroscopic system \eqref{eq:macro_eq1}--\eqref{eq:macro_eq3} can be obtained as a relaxation limit for $\eta \to 0$ of the following conservative system with source term:
\begin{align}
    \label{eq:Relaxation1}
    &\p_t \rho^\eta + c_1 \nabla_x \cdot \left(\rho^\eta\Omega^\eta \right) = 0\,,\\
    \label{eq:Relaxation2}&\p_t \left(\rho^\eta \omega^\eta \right) + c_1 \nabla_x \cdot \left(\rho^\eta \omega^\eta \Omega^\eta \right) = 0\,,\\
    &\label{eq:Relaxation3} \p_t \left(\rho^\eta \Omega^\eta \right) + \frac{K_2}{K_1} \nabla_x \cdot \left( \rho^\eta \Omega^\eta \otimes \Omega^\eta\right) +\nabla_x \rho^\eta -\omega^\eta(\Omega^\eta) ^\perp \rho^\eta = \frac{\rho^\eta}{\eta}\left( 1- |\Omega^\eta|^2\right)\Omega^\eta \,,
\end{align}
where $C = \frac{k_\theta}{\alpha^2}$. We first solve the conservative part using a custom Roe finite volume scheme and then treat the relaxation and rotational source terms by a splitting method \cite{motsch2011numerical}. The resulting scheme relies on the following steps:
\begin{enumerate}
    \item We first solve the conservative part of the relaxation system 
    \begin{align}
    \label{eq:MacroConservative1}
    &\p_t \rho  + c_1\nabla_x \cdot \left(\rho\, \Omega  \right) = 0\,,\\
    \label{eq:MacroConservative2}&\p_t (\rho\,  \omega\,) + c_1\nabla_x \cdot (\rho \, \omega \, \Omega  ) = 0\,,\\
    &\label{eq:Macroconservative3} \p_t \left(\rho \, \Omega  \right) + \frac{K_2}{K_1} \nabla_x \cdot \left( \rho\,  \Omega  \otimes \Omega  +  \rho  \mathbb{I}_2\right)  = 0 \,,
    \end{align}
     which corresponds to a compressible Euler-type system in two dimensions. We solve it by dimensional splitting into successive one-dimensional Euler systems, each discretized with a custom Roe scheme with a Roe matrix computed
    following \cite{leveque2002finite} page 156.
    \item As shown in \cite{motsch2011numerical}, in the limit for \(\eta \to 0\), the relaxation part 
    \begin{align}
    \label{eq:MacroRelaxation1}
    &\p_t \rho^\eta  = 0\,,\\
    \label{eq:MacroRelaxation2}&\p_t \left(\rho^\eta \omega^\eta \right) = 0\,,\\
    &\label{eq:MacroRelaxation3} \p_t \left(\rho^\eta \Omega^\eta \right)  = \frac{\rho^\eta}{\eta}\left( 1- |\Omega^\eta|^2\right)\Omega^\eta \,,
\end{align}
reduces to a normalization of \(\Omega\).
\item Finally, the source term
\begin{align}
    \label{eq:MacroSource1}
    &\p_t \rho  = 0\,,\\
    \label{eq:MacroSource2}
    &\p_t (\rho\, \omega) = 0\,,\\
    \label{eq:MacroSource3}
    &\p_t \Omega = \omega \, \Omega^\perp\,,
\end{align}
can be solved explicitly in dimension two. Writing
\(
\Omega(t) = (\cos \theta(t), \sin \theta(t))^{ T},
\)
the last equation simply gives
\(
\theta(t) = \theta_0 + \omega\,t\,.
\)
\end{enumerate}
The implementation of our scheme is freely available at \cite{carmela_moschella_2025_17945342}
and is a straightforward adaptation of the code by Antoine Diez in \cite{Diez_SOHjlj_2025}, 
originally developed for the Self-Organized Hydrodynamics (SOH) model. A previous code for the SOH model in Fortran by Sebastien Motsch can also be found in \cite{smotsch_vicsek_macro_2025}.

Simulations of the macroscopic Vicsek--Kuramoto  model are performed on a square periodic domain \( [0, L_x] \times [0, L_y] \), with \( L_x = L_y = 1 \). We fix the time step to \( \Delta t = 0.001 \) and integrate the system up to final time \( t_{\text{end}} = 100 \), while \( \Delta x = \Delta y = 0.005 \). The boundary conditions in both directions are taken to be periodic. The initial conditions for the macroscopic simulations are qualitatively comparable with those used in the microscopic setting. Specifically, the density field \( \rho(x,0) \) is initialized as a spatially homogeneous profile with small random perturbations, while the angular velocity field \( \omega(x,0) \) is prescribed with a non-zero spatial mean to replicate the asymmetric distribution of angular velocities employed in the simulations of the microscopic system. The mean orientation field \( \Omega(x,0) \) is sampled uniformly from the unit circle. We choose \( {k_\theta}/{\alpha^2} = 8 \), as justified in remark \ref{rem:kalpha}.

\subsubsection{Patterns observed}

We conducted simulations across a broad range of parameter values and consistently observed synchronized dynamics as the long-time behavior. In some runs, the system exhibited transient patterns — such as rotating clusters — before converging to the stationary synchronized state. Representative examples are shown in Fig. \ref{fig:macro_simulation}, and videos corresponding to the snapshots in that figure are provided in Appendix \ref{appendix:videos} (Videos 5 and 6). These results suggest that the macroscopic equations predominantly capture synchronized behavior in the long-time limit. In the next section, we discuss potential explanations for this observation.

We also checked numerically that,  as claimed in Cor. \ref{cor:constant_initial_data}, constant initial data produces a solution with synchronized behavior.

\begin{figure}[htbp]
    \centering
    \hspace{1em}
    \begin{subfigure}[b]{0.40\textwidth}
        \includegraphics[width=\textwidth]{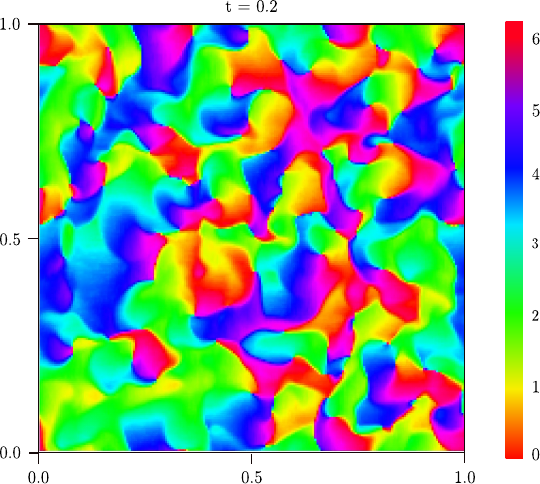}
        \caption{}
    \end{subfigure}
    \hfill
    \begin{subfigure}[b]{0.40\textwidth}
        \includegraphics[width=\textwidth]{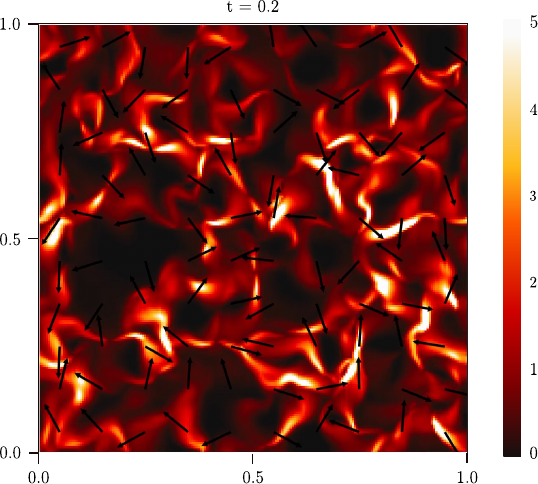}
        \caption{}
    \end{subfigure}
    \hspace{1em}
    
    \vspace{0.5em}
    \hspace{1em}
    \begin{subfigure}[b]{0.40\textwidth}
        \includegraphics[width=\textwidth]{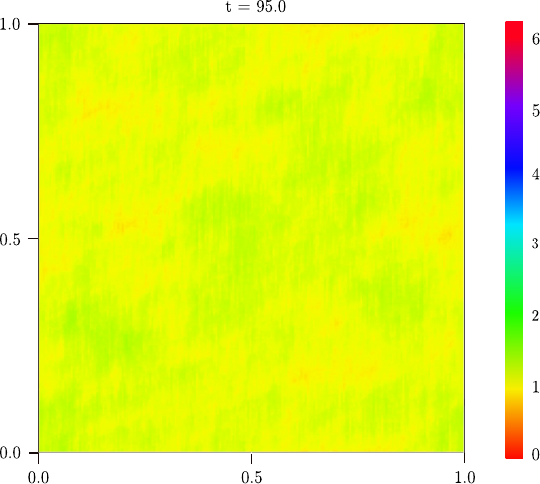}
        \caption{}
    \end{subfigure}
    \hfill
    \begin{subfigure}[b]{0.40\textwidth}
        \includegraphics[width=\textwidth]{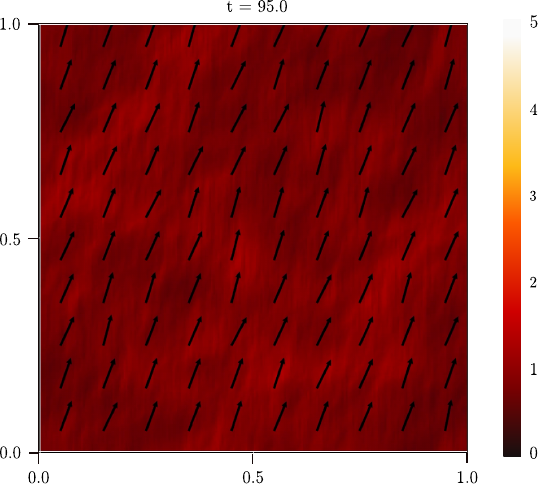}
        \caption{}
    \end{subfigure}
    \hspace{1em}
    
\caption{
Macroscopic simulation of the Vicsek--Kuramoto system with \( k_\theta / \alpha^2 = 8 \) starting from random initial data. 
\textbf{(a)} Snapshot at early time \( t = 0.20 \). The color scale represents the density \( \rho(x,t) \) and it shows the formation of localized rotating clusters. These are transient structures similar to those observed in the microscopic simulations with small \( k_\theta \) and \( k_\omega \), corresponding to weak orientational and angular velocity alignment. The video of the simulation is included in Appendix~\ref{appendix:videos} (Video~5)
\textbf{(b)} Same configuration as in (a), with the color scale representing the average orientation angle of each cell grid between $[-\pi, \pi]$. Arrows indicate the local mean orientation field $\Omega(x,t)$. The video of the simulation is included in Appendix~\ref{appendix:videos} (Video~6)
\textbf{(c)} Snapshot of Video 5 at time \( t = 95.0 \). The density \( \rho \) becomes nearly uniform in space, and the system reaches a rotating regime driven by a non-zero mean average angular velocity.
\textbf{(d)} Snapshot of Video~6 at time t =95, with colorbar indicating the angle. The global coherent rotation of the direction field confirms the emergence of a synchronised state, in qualitative agreement with the behavior observed in the microscopic simulations of Video 4.
}
    \label{fig:macro_simulation}
\end{figure}

\subsection{Summary and interpretation of the results}
The numerical simulations presented in Sections \ref{sec:simulation_micro} and \ref{sec:simulation_macro} provide a qualitative characterization of the patterns that arise in both the microscopic and macroscopic models. The results indicate that the interplay between orientation alignment and angular velocity alignment is the key mechanism responsible for the emergence of distinct collective behaviors in the IBM.
In the regime where both the alignment in angular velocity and direction are weak, the system self-organizes in localized rotating clusters. As the alignment strength in orientation becomes dominant over the one in angular velocity, the system exhibits tighter spatial aggregation, with clusters becoming more compact and localized. Conversely, when angular velocity alignment dominates and the orientation alignment remains weak, the particle distribution tends toward a spatially homogeneous configuration with minimal clustering. 

Among the various regimes explored, the synchronised state emerges as the most robust and persistent collective pattern. This regime is typically observed when both alignment strengths are moderate to large. In contrast, traveling waves in the orientation field arise only within a narrow range of parameters, specifically when both the orientation and angular velocity alignment strengths lie in a low-to-intermediate regime. In fact, the macroscopic equations are derived under a hydrodynamic scaling in which both alignment interactions become large as $\varepsilon \to 0$. This could explain why traveling waves do not appear in the macroscopic simulations, which may lie outside the range of validity of the macroscopic approximation considered in this work. 
Accordingly, while the macroscopic model may initially exhibit localized rotating structures reminiscent of the microscopic clustering regime, this behavior persists only for a few time iterations before being rapidly averaged out. The system then tends toward a globally synchronized rotational state.

Notably, the parameters $k_\omega$ and $\beta^2$ — whose ratio plays a crucial role in determining patterns in the IBM — do not appear explicitly in the macroscopic equations. As a result, the continuum model effectively removes this degree of freedom, which limits its ability to reproduce the full spectrum of microscopic behaviors. This is not unusual: the averaging intrinsic to continuum descriptions smooths out fine-scale structures and stochastic fluctuations in the macroscopic limit. Similar limitations are observed in other individual-based systems: phenomena such as clustering, bistability, or symmetry breaking—clearly visible in the IBM—may be absent from the corresponding macroscopic formulation~\cite{carrillo2009double}. On top of this, some patterns may be driven by phase-transition behaviour, as in the Vicsek model~\cite{vicsek1995}. In our treatment of the IBM, we deliberately excluded such phase transitions; however, alternative modelling choices for Vicsek-type systems can retain them, as demonstrated in~\cite{degond2013phase}.

\section{Conclusions and perspectives}
In this work, we introduced and analyzed a Vicsek--Kuramoto (VK) model for self-propelled agents with coupled alignment of orientation and angular velocity, derive its mean-field (Fokker--Planck) kinetic description, and obtain a closed hydrodynamic (Euler-type) system via Generalized Collisional Invariants for the agent density, mean orientation, and average angular velocity. We complement the analysis with microscopic and macroscopic simulations, revealing regimes where the macroscopic model captures large-scale behaviors for the IBM, thereby clarifying its domain of validity. In particular, the IBM presents various types of patterns, including rotating clusters, travelling waves for the orientation, and synchronised behaviour where all particles rotate with the same frequency. Of all these patterns, the synchronised behaviour seems to be the prevalent one in the macroscopic regime. 

The model establishes a connection between synchronization and collective motion in active matter, and it—along with suitable adaptations—can be employed to investigate patterns observed in biological suspensions \cite{weaksync,sumino2012vortex,riedel2005self}. Future directions include extending the framework to alternative alignment mechanisms (e.g., nematic or apolar interactions) \cite{degond2020nematic,degond2018age} and incorporating particle-shape effects (e.g., elongated agents) by modeling steric interactions through anisotropic repulsion potentials~\cite{merino2025macroscopic}.

\section*{Acknowledgments}
The authors would like to thank Antoine Diez and Steffen Plunder for wonderful and fruitful discussions.\\

The work of SMA was funded in part by the Austrian Science Fund (FWF) project \href{https://www.fwf.ac.at/en/research-radar/10.55776/F65}{10.55776/F65}
and in part by the Vienna Science and Technology Fund (WWTF) \href{https://www.wwtf.at/funding/programmes/vrg/VRG17-014/index.php?lang=EN}{[10.47379/VRG17014]}. The work of CM was funded by the Austrian Science Fund (FWF), project \href{https://www.fwf.ac.at/forschungsradar/10.55776/W1261}{10.55776/W1261}.

\printbibliography

\appendix

\section{Auxiliary computations}

\subsection{Proof of equality \eqref{eq:integralc1}}
\label{sec:proof_integral}
We consider
\begin{equation*}
   \int_{-\pi}^{\pi}
        \vec\tau(\theta)\,
        \frac{e^{\kappa\cos(\theta-\bar\theta_0)}}{2\pi I_{0}(\kappa)}
        \,d\theta =  \int_{-\pi}^{\pi}
        (\cos \theta, \sin \theta)\,
        \frac{e^{\kappa\cos(\theta-\bar\theta_0)}}{2\pi I_{0}(\kappa)}
        \,d\theta\,,
\end{equation*}
and we
set \(\Gamma=\theta-\bar\theta_0\), obtaining
\[
\int_{-\pi - \theta_0}^{\pi - \theta_0} \left(\cos(\Gamma + \theta_0), \sin(\Gamma + \theta_0)\right) \frac{e^{k \cos \Gamma}}{2\pi I_{0}(\kappa)} \, d\Gamma = \int_{-\pi}^{\pi} \left(\cos(\Gamma + \theta_0), \sin(\Gamma + \theta_0)\right) \frac{e^{k \cos \Gamma}}{2\pi I_{0}(\kappa)} \, d\Gamma\,,
\]
where in the last equality we have used the fact that each term is $2\pi-$periodic. We now use 
$$\left(\cos(\Gamma + \theta_0), \sin(\Gamma + \theta_0)\right)^T  = \begin{pmatrix}
\cos \Gamma \cos \theta_0 - \sin \Gamma \sin \theta_0 \\
\sin \Gamma \cos \theta_0 + \cos \Gamma \sin \theta_0
\end{pmatrix}\,,$$
 Since
\(\sin\Gamma\,e^{\kappa\cos\Gamma}\) is odd, its integral vanishes, while the $\cos\Gamma$ part yields
\[
\int_{-\pi}^{\pi}
      \cos\Gamma\,
      \frac{e^{\kappa\cos\Gamma}}{2\pi I_{0}(\kappa)}\,d\Gamma
      \;=\;
      \frac{I_{1}(\kappa)}{I_{0}(\kappa)} = c_1.
\]
We can the conclude that 
\[\int_{-\pi}^{\pi}
        \vec\tau(\theta)\,
        \frac{e^{\kappa\cos(\theta-\bar\theta_0)}}{2\pi I_{0}(\kappa)}
        \,d\theta = c_1 (\cos \theta_0, \sin \theta_0) =c_1 \Omega^0\,, \]
where $\Omega^0= (\cos \theta_0, \sin \theta_0).$

\section{List of supplementary videos}\label{appendix:videos}
The supplementary videos can be found at the following link:
\begin{center}
\url{https://figshare.com/account/home#/projects/252416}
\end{center}
\subsection{Particle simulations}
These videos illustrate the phenomena described in Section \ref{sec:phase_diagrams}.

\

\noindent \textbf{Video 1} Video corresponding to Figure~\ref{fig:phase_diagram}, with parameters described in Table~\ref{tab:parameters_micro}, and \( k_\theta = k_\omega = 1 \), \( R = 2 \). In this regime, the system self-organizes into rotating, cluster-like aggregates. Although all clusters share the same angular velocity (in absolute value), their rotational phases differ, leading to a configuration of multiple rotating clusters without phase synchronization.

\

\noindent \textbf{Video 2} Video corresponding to Figure~\ref{fig:traveling_micro}, with parameters described in Table~\ref{tab:parameters_micro}, and \( k_\theta = 21 \), \( k_\omega = 81 \), and \( R = 2 \). In this regime, the system exhibits a traveling wave configuration in the orientation field while maintaining a nearly uniform spatial distribution.

\

\noindent \textbf{Video 3} Video corresponding to Figure \ref{fig:synchronised}, with parameters described in Table \ref{tab:parameters_micro}, and $k_\theta= k_\omega = 71$, and $R=2$. In this regime, the particles are uniformly distributed in space and exhibit collective rotation, moving as a unique rigid body.

\

\noindent \textbf{Video 4.} Time evolution of the microscopic Vicsek--Kuramoto system used for qualitative comparison with the macroscopic simulation in Figure~\ref{fig:macro_simulation}. The simulation is performed with \( N = 2\cdot10^4 \) particles, alignment strengths \( k_\theta = 25.0 \), \( k_\omega = 50.0 \), noise intensities \( \sqrt{2 \alpha^2} = 2.5 \), \( \sqrt{2 \beta^2} = 2.5 \), self-propulsion speed \( c = 0.1 \), interaction radius \( R = 0.04 \), and domain size \( L = 1 \). In this regime, the system exhibits the same synchronised behavior observed in Video 3. This behavior qualitatively matches the macroscopic dynamics observed when \(k_\theta / \alpha^2 = 8 \).

\subsection{Macroscopic simulations}

\noindent \textbf{Video 5.} Time evolution of the macroscopic Vicsek--Kuramoto system simulated with  \( k_\theta / \alpha^2 = 8 \), corresponding to the results shown in Figure~\ref{fig:macro_simulation}. The color scale represents the density \( \rho(x,t) \) of the system.

\medskip

\noindent \textbf{Video 6.} Same simulation set up as in Video~5. The color scale now represents the average orientation angle \( \theta(x,t) \) between $[-\pi, \pi]$.

\end{document}